\documentclass[journal]{IEEEtran}

\usepackage[compress]{cite}
\usepackage{stmaryrd}
\usepackage{mathrsfs}
\usepackage{amsfonts}
\usepackage{amsmath}
\usepackage{cases}
\usepackage{txfonts}
\usepackage{amssymb}
\usepackage{epsfig}
\usepackage{pdfpages}
\usepackage{algorithm}
\usepackage{algpseudocode}
\usepackage{slashbox}
\usepackage{multirow}

\newcommand{\fat}[1]{\mbox{\boldmath$#1$}}

\newtheorem{theorem}{Theorem}

\newtheorem{definition}{Definition}

\newcommand{\myQED}{\mbox{}\hfill{$\Box$}}

\begin{document}

\title{Performance Limit and Coding Schemes for Resistive Random-Access Memory Channels}

\author{Guanghui Song, \IEEEmembership{Member, IEEE},
        Kui Cai, \IEEEmembership{Senior Member, IEEE}, Xingwei Zhong\\ Jiang Yu, and Jun Cheng, \IEEEmembership{Member, IEEE}
}

\maketitle

\begin{abstract}
Resistive random-access memory (ReRAM) is a promising candidate for the next generation non-volatile memory technology due to its simple read/write operations and high storage density.
However, its crossbar array structure causes a severe interference effect known as the ``sneak path."
In this paper, we propose channel coding techniques that can mitigate both the sneak-path interference and the channel noise. The main challenge is that the sneak-path interference is data-dependent, and also correlated within a memory array, and hence the conventional error correction coding scheme will be inadequate.
In this work, we propose an across-array coding strategy that assigns a codeword to multiple independent memory arrays, and exploit a real-time channel
estimation scheme to estimate the instantaneous status
of the ReRAM channel. Since the coded bits from different
arrays experience independent channels, a ``diversity" gain can be
obtained during decoding, and when the codeword is adequately
distributed over different memory arrays, the code actually performs as that over
an uncorrelated channel. By performing decoding based on the scheme of treating-interference-as-noise (TIN), the ReRAM channel over different memory arrays is equivalent to a block varying channel we defined, for which we propose both the capacity bounds and a coding scheme. The proposed coding scheme consists of
a serial concatenation of an optimized error correction code with a data shaper, which enables the ReRAM system to achieve a near capacity limit  storage efficiency.
\end{abstract}


\section{Introduction}
Resistive random-access memory (ReRAM) is an emerging non-volatile memory technology that changes the resistance value of a memristor to represent two states of the binary user data: the High-Resistance State corresponding to logic 0 while the Low-Resistance State corresponding to logic 1. The memristor cell is positioned on each row-column intersection of a crossbar structure, which offers a huge density advantage for ReRAM systems \cite{Strukov}.

When a cell in a memory array is read, voltage is applied to the memristor cell, and the current flows through the memristor and senses the resistance value.  If the memristor is detected with a High-Resistance State, the bit is decided to be a `0'; if it is detected with a Low-Resistance State, the bit is determined to be a `1'. A fundamental drawback of the ReRAM crossbar array is the sneak-path problem \cite{Zidan}.  Sneak paths are undesirable paths in parallel to the selected cell for reading. The current goes through the sneak paths and degrades the measured resistance value. Sneak paths are detrimental when a cell with a High-Resistance State is read because the resistance degradation may lead to an erroneous decision.

In the literature, several works \cite{Yuval,Ben,CZH} tackled the sneak-path problem by using information and communication theoretical frameworks. In particular, Y. Cassuto \emph{et al.} \cite{Yuval} studied the maximum storage efficiency when the constrained codes are employed to completely avoid sneak paths.  This method incurs a high code rate loss, especially when the array size is large. Y. Cassuto \emph{et al.}  proved that as the array size approaches infinity, the storage information rate approaches 0 in order to achieve a sneak-path free channel. On the other hand, a commonly used method to eliminate the sneak paths is to introduce a cell selector in series to each array cell. However, these selectors are also prone to failure due to the imperfections of the memory fabrication and maintenance process, leading to the reoccurrence of the sneak paths \cite{Ben}\cite{CZH}. Y. Ben-Hur \cite{Ben} and Zehui \emph{et al.} \cite{CZH} considered ReRAM systems with imperfect selectors which fail with a certain probability. They built a probabilistic sneak-path model and developed the corresponding data detection schemes. A main challenge for the ReRAM channel is that the sneak-path induced interference to the channel is data-dependent. Moreover, the sneak-path interference is also correlated between different locations of the crossbar array. Previous work  \cite{Ben} developed single-cell data detection schemes that detect the data for each memory cell independently.  More sophisticated joint-cell data detection schemes were developed in \cite{CZH} by introducing pilot cells.
However, the probabilistic model in \cite{CZH} becomes too complex when the array size is getting large. No error correction code (ECC) was employed in previous research works \cite{Ben}\cite{CZH}.

According to the information theory, an efficient way to achieve reliable data storage is to apply ECC to the system.  Such a design should not be a straightforward extension of the conventional ECC designed for the symbol-wise independent and identically distributed  (i.i.d.) channels. Other than the channel noise, the code must overcome the sneak-path interference which is data-dependent and also correlated within the crossbar array.

In this paper, we propose efficient coding and decoding schemes for ReRAM channels. To reduce the correlation of the sneak-path interference within a codeword, we propose an across-array coding strategy which spreads a codeword to multiple independent memory
arrays, and also exploit a real-time channel estimation scheme to obtain the channel status of each memory array.  Since the coded bits from different memory arrays experience independent channels, the overall channel will be averaged and a ``diversity" gain will be obtained during decoding. In this way, we can design the coding scheme over a symbol-wise i.i.d. channel.  By further applying the treating-interference-as-noise (TIN) decoding, the ReRAM channel is equivalent to a block-varying channel whose status does not depend on the input data, based on which we derive the lower and upper bounds of its channel capacity and propose a coding scheme.  The proposed coding scheme consists of an outer irregular repeat-accumulate (IRA) code being concatenated with an inner data shaper, which is used to change the input data distribution so as to achieve the maximum information rate. A low-complexity joint message-passing decoding for the IRA code and the data shaper is developed based on the state-of-the-art sparse-graph coding theory. With an optimized IRA code, our ReRAM system achieves a near capacity limit storage efficiency.

The rest of this paper is organized as follows. In Section~\ref{sec:model}, we present the ReRAM channel model and describe the data-dependent feature of the sneak-path interference.
The  across-array coding strategy and the capacity bounds for the ReRAM channel are proposed in Section~\ref{sec:capacity}. In Section~\ref{sec:coding}, we propose a coding scheme for ReRAM channels and present both numerical and simulation results.  In Section~\ref{sec:conclusion}, we conclude the paper.
\section{ReRAM Channel Model}\label{sec:model}
Consider an $m\times n$ crossbar memory array.  The memristor that lies at the intersection of row $i$ and column $j$ denotes memory cell $(i, j)$.
Each array is able to store an $m\times n$ binary data matrix $\fat{X}=[x_{i,j}]_{m\times n}$, where bit $x_{i,j}\in\{0, 1\}$  is stored in memory cell $(i, j), i\in\{1,..., m\}, j\in\{1,..., n\}$. During the writing process, each bit is written into the memory cell by changing the resistance value of the memristor, i.e., for a logical ``0" bit, the cell is changed to a high resistance of $R_0$, referred to as the High-Resistance State, and for a logical ``1" bit, it is changed to a low resistance of $R_1$, referred to as the Low-Resistance State.

\begin{figure}[t]
\includegraphics[width=
3.5 in]{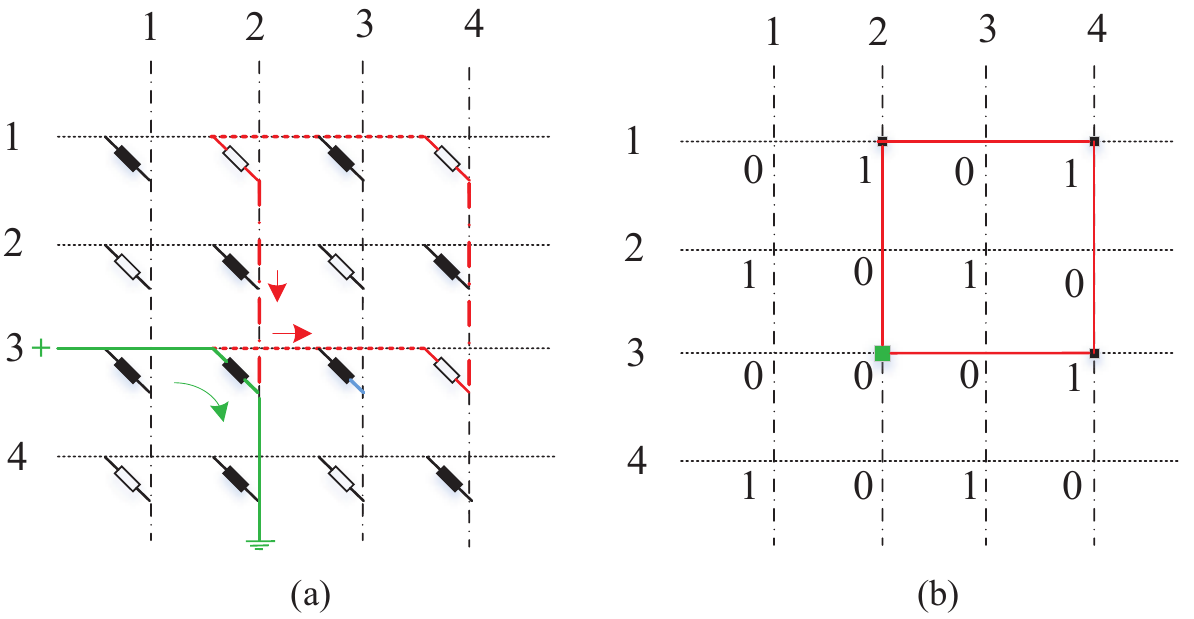}
\centering
\caption{(a) Example of a $4\times 4$ memory array.  (b) Corresponding logical values of memory array. $(3, 2)$ is a target cell for reading. Voltage is applied to memristor cell $(3, 2)$ and green line is the desired path for resistance measuring. However, $(3, 2)\rightarrow(3, 4)\rightarrow(1, 4)\rightarrow(1, 2)\rightarrow(3, 2)$ forms a sneak path (red line) in parallel of target cell $(3, 2)$ that degrades the measured resistance value. Arrows show current flow directions. Note that the sneak path produces a reverse current across cell $(1, 4)$.} \label{fig:SPmodel}
\end{figure}

During the reading process, the data bit can be detected by measuring the resistance value of the corresponding cell. If it is in the High-Resistance State, the bit is identified as a `0'; if it is in the Low-Resistance State, the bit is identified as a `1'. However, due to the existence of the sneak-path interference, as well as a mix of other noises which can be modeled as an additive Gaussian noise, the memory reading becomes unreliable \cite{Yuval,Ben,CZH}. When the cell $(i, j)$ in a memory array is read, certain voltage is applied to the target cell to measure its resistance. A sneak path is defined as a closed path that originates from and returns to location $(i, j)$ while traversing logical-1 cells through alternating vertical and horizontal steps. An example is shown in Fig.~\ref{fig:SPmodel}, where $(3, 2)$ is a target cell for reading and the green line shows the desired path to measure the resistance. However, $(3, 2)\rightarrow(3, 4)\rightarrow(1, 4)\rightarrow(1, 2)\rightarrow(3, 2)$ forms a sneak path (red line) in parallel of the selected cell $(3, 2)$.
Since sneak paths always degrade the measured resistance value, they actually benefit the data detection when a cell in a Low-Resistance State (logic 1) is read. The detrimental effect only occurs when a High-Resistance State cell (logic 0) is read, making it more vulnerable to noise.
 In this paper, we only consider the sneak path when a High-Resistance State cell is read.

The most popular method to mitigate the sneak-path interference is to introduce a cell selector in series to each array cell. A cell selector is an electrical device that allows current to flow only in one direction across the cell. Since sneak paths inherently produce reverse currents in at least one of the cells along the parallel path, the cell selector can completely eliminate sneak paths from the entire array.  In this paper we follow the previous work \cite{Ben,CZH} and consider the 1D1R structure, even though our proposed approaches can be extended to other structures as well. Although cell selectors can effectively eliminate sneak paths, they are also prone to failures due to the imperfections in the production or the maintenance of memory array, leading to the reoccurrence of the sneak paths. Following previous work \cite{Ben,CZH}, we assume that the selectors in a memory array fail i.i.d. with probability $p_f$. However, our work is based on the fact that once a selector fails it will by no means recover, and hence the locations of the failed selectors are fixed during the reading of the whole array. This is different from the assumption made by the previous work \cite{Ben,CZH} that the locations of the failure cells vary randomly in the crossbar array during the reading of each cell. Although these two assumptions may not affect the detection performance significantly for the uncoded ReRAM systems, they are fundamentally different for the coded systems. The previous assumption \cite{Ben,CZH} actually leads to a near i.i.d. model for the sneak-path interference of each cell. In this work, the sneak-path events for the cells at different locations in the same array are highly correlated, which is much more difficult to be tackled by the channel coding scheme.

 We define a sneak-path event indicator $e_{i, j}$ for cell $(i, j)$ to be a Boolean variable with the value $e_{i, j}=1$ if the cell $(i, j)$ is affected by sneak paths, otherwise, $e_{i, j}=0$. According to the previous work \cite{CZH}, sneak-path events occur during the reading of cell $(i, j)$ and lead to $e_{i, j}=1$ if and only if the following three conditions are satisfied:

\emph{[Sneak-Path Condition:]}

1) The cell $(i, j)$ is in a High-Resistance State, i.e., $x_{i, j}=0$.

2) There exists at least one combination of $k\in\{1,...,m\}, \ell\in\{1,...,n\}$ that induces a sneak path, defined by
\begin{equation}
x_{i, \ell}=x_{k, \ell}=x_{k, j}=1.
\end{equation}

3) The selector at the diagonal cell $(k, \ell)$ fails. Due to the circuit structure of the crossbar array, cells $(i,l)$ and $(k,j)$ will conduct current in the forward direction and not be affected by their selectors.  Only when the selector of the cell $(k,l)$ is faulty, a sneak path will be formed.

The above Sneak-Path Condition definition limits the sneak path to length of 3, i.e., traversing three cells \cite{CZH}. More sophisticated cases of the sneak paths were considered in \cite{Ben}. The principle of our work can be extended to those cases.

Building on the above Sneak-Path Condition, we define a

\emph{[ReRAM Channel:]}

 Let $\fat{X}=[x_{i,j}]_{m\times n}$ be the stored data array and $\fat{Y}=[y_{i,j}]_{m\times n}$ be the corresponding readback signal for the crossbar array. Let $\mathcal{R}$ be the set of real numbers.
An ReRAM channel is a channel with input $\fat{X}\in \{0, 1\}^{m\times n}$ and output $\fat{Y}\in \mathcal{R}^{m\times n}$:
\begin{equation}
y_{i,j}=\begin{cases} (\frac{1}{R_0}+\frac{e_{i,j}}{R_s})^{-1}+\eta_{i,j}&\mbox{if $x_{i,j}=0$}\\ R_1+\eta_{i,j}&\mbox{if $x_{i,j}=1$} \end{cases}
\label{eq:ReRAMchannel}
\end{equation}
where $R_s$ is the parasitic resistance value brought by sneak paths. Here $\eta_{i,j}\sim \mathcal{N}(0, \sigma^2), i=1,...,m, j=1,...,n$ is an additive white Gaussian noise (AWGN) with mean 0 and variance $\sigma^2$ \cite{Ben}. It is used to model a mix of various noises of the ReRAM system.

\begin{figure}[t]
\includegraphics[width=
3.2 in]{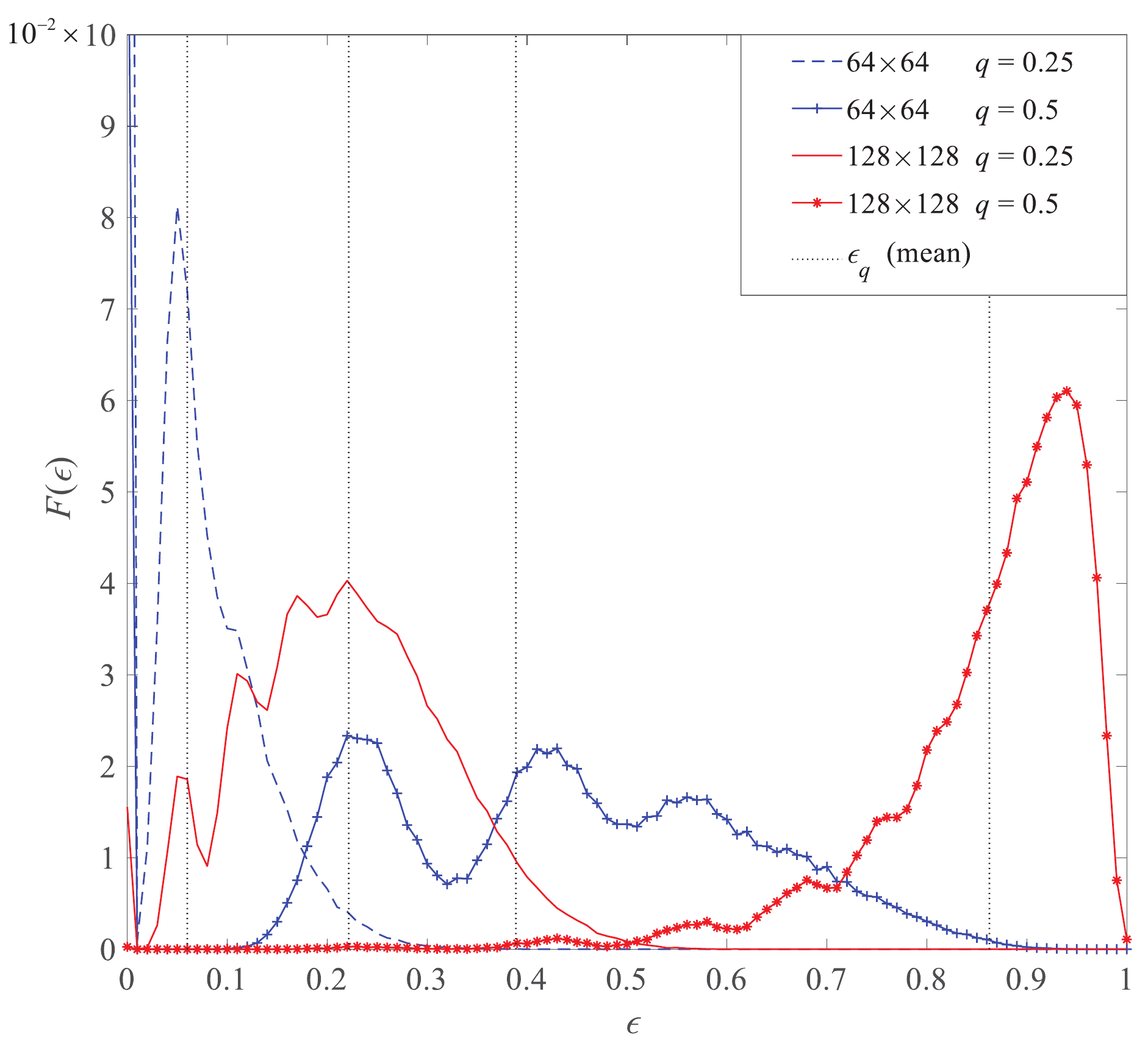}
\centering
\caption{PMFs of sneak-path rate within single memory array simulated for array sizes $m\times n=64\times64, 128\times128$ and input distributions with $q=0.25, 0.5$. The mean values of sneak-path rates for the four cases, indicated by the dashed lines, are $\epsilon_q=0.06, 0.3888, 0.2216$, and $0.8626$, respectively.} \label{fig:SP_rate}
\end{figure}

The fundamental problem of the ReRAM channel is to recover the stored data array $\fat{X}$ based on readback signal $\fat{Y}$ in the presence of sneak-path interference $[e_{i,j}]_{m\times n}$ and Gaussian noise $[\eta_{i,j}]_{m\times n}$. The ReRAM channel $\{0, 1\}^{m\times n}\rightarrow \mathcal{R}^{m\times n}$ with input and output size $m\times n$ actually consists of $mn$ symbol-wise channels with $\{0, 1\}\rightarrow \mathcal{R}$. Since the sneak-path indicator $e_{i,j}$ of each target cell is related to the entire data array, these $mn$ symbol-wise channels are data-dependent and correlated.

The ReRAM channel is also asymmetrical, whose channel status (sneak-path occurring probability) is highly related to the channel input distribution. We define the input distribution as i.i.d. Bernoulli $(q)$, i.e., $\textrm{Pr}(x_{i,j}=1)=q$ and $\textrm{Pr}(x_{i,j}=0)=1-q$ for $i=1,..., m, j=1,..., n$.

For a fixed input distribution, we investigate the fraction of sneak-path affected cells in the crossbar array and define a sneak-path rate over the array as $\frac{\sum_{i=1}^m\sum_{j=1}^ne_{i,j}}{mn(1-q)}$. Its mean value is derived as a function of $q$:
\begin{eqnarray}
\epsilon_q\!\!\!\!\!\!\!\!&&\overset{\Delta}{=}E\left[\frac{\sum_{i=1}^m\sum_{j=1}^ne_{i,j}}{mn(1-q)}\right]\label{eq:SP-rate}\\
\!\!\!\!\!\!\!\!&&=\textrm{Pr}(e_{i,j}=1| x_{i,j}=0)\nonumber\\
\!\!\!\!\!\!\!\!&&=1-
\sum_{u = 0}^{m - 1} \sum_{v = 0}^{n - 1}\binom{m-1}{u} \binom{n-1}{v}q^{u + v}( 1 - q )^{m - 1 - u + n - 1 - v}\nonumber\\
\!\!\!\!\!\!\!\!&&\ \ \ \ \ \ \ \ \ \ \ \ \ \ \ \times(1 - p_fq)^{uv}.\label{eq:SP-rate-cal}
\end{eqnarray}
When $m$ or $n$ is large,  (\ref{eq:SP-rate-cal}) is difficult to calculate.
However, when $p_fq$ is small, using the Taylor expansion $(1 - p_fq)^{uv}\approx 1 -uv p_fq+\alpha\binom{uv}{2}p_f^2q^2$, we can approximately calculate (\ref{eq:SP-rate-cal}):
\begin{eqnarray}
\epsilon_q\!\!\!\!\!\!\!\!&&\approx1-
\sum_{u = 0}^{m - 1} \sum_{v = 0}^{n - 1}\binom{m-1}{u} \binom{n-1}{v}q^{u + v}( 1 - q )^{m - 1 - u + n - 1 - v}\nonumber\\
\!\!\!\!\!\!\!\!&&\ \ \ \ \ \ \ \ \ \ \ \ \  \ \ \times\left(1 -uv p_fq+\alpha\binom{uv}{2}p_f^2q^2\right)\label{eq:talor}\\
\!\!\!\!\!\!\!\!&&=(m-1)(n-1)p_fq^3-\alpha\left(2q\binom{m-1}{2}\binom{n-1}{2}\right.\nonumber\\
\!\!\!\!\!\!\!\!&&\ \ \ \ \ \ \ \ \ \ \ \ \left.+(n-1)\binom{m-1}{2}+(m-1)\binom{n-1}{2}\right)p_f^2q^5
\end{eqnarray}
 where $\alpha$ is a balance factor for the last term of the Taylor expansion.  Here a good setting for $\alpha$ is $0.8$.

We define the probability mess function (PMF) of the sneak-path rate as
\begin{equation}
F(\epsilon)=\textrm{Pr}\left(\frac{\sum_{i=1}^m\sum_{j=1}^ne_{i,j}}{mn(1-q)}=\epsilon\right).
 \end{equation}
 For a memory array size of $m\times n=64\times64, 128\times128$ and input distributions with $q=0.25, 0.5$, we simulate $F(\epsilon)$ and the results are illustrated by Fig.~\ref{fig:SP_rate}. In particular, we generate a large amount of input data arrays, compute the sneak-path rate of each array, and obtain the PMF statistically. In the simulations as well as the numerical results of this paper, we assume the selectors fail i.i.d. with probability $p_f=10^{-3}$.  Fig.~\ref{fig:SP_rate} shows that a larger value of $q$, or a larger array size, leads to higher sneak-path rates, i.e., worse channels. The values of the sneak-path rate are quite diverse for different input data patterns since the PMF spreads in a large range over the abscissa, which indicates that the channel varies significantly for different input data patterns. This is because the occurrence of sneak-path events depends on the input data pattern.
This creates a big challenge for designing the coding scheme for the ReRAM channels since the code directly designed based on the average sneak-path rate of $\epsilon_q=\sum_{\epsilon}\epsilon F(\epsilon)$ will be inadequate.

\begin{figure*}[t]
\includegraphics[width=
5.8 in]{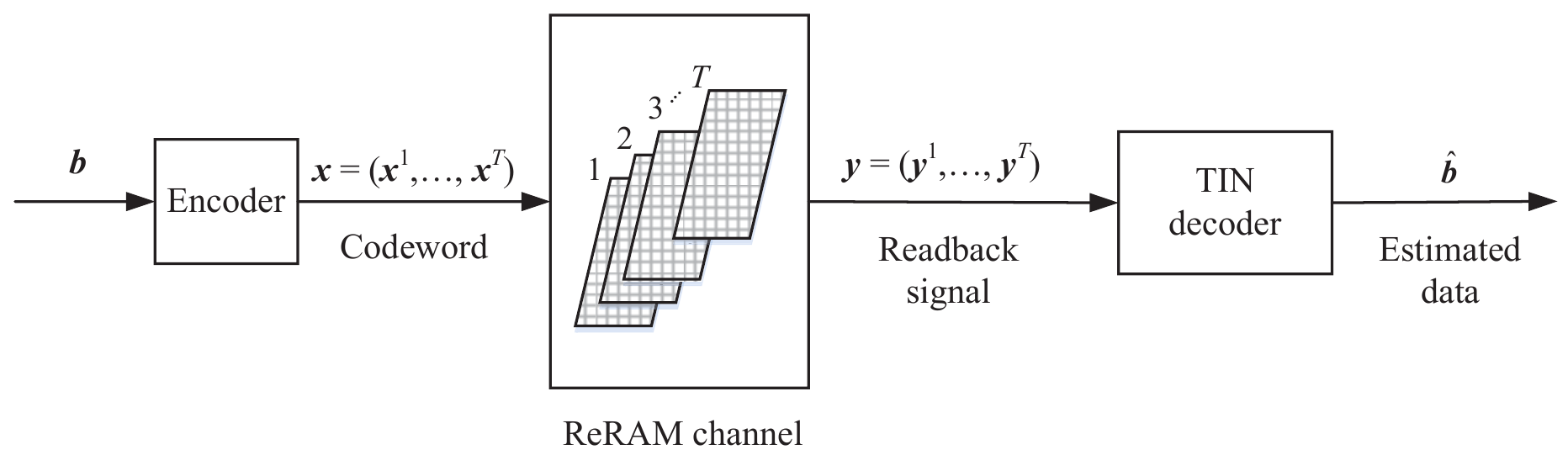}
\centering
\caption{Across-array coding strategy for ReRAM.} \label{fig:Coding}
\end{figure*}

\section{Across-Array Coding Strategy and Channel Capacity Bounds}\label{sec:capacity}
To mitigate the variability of the ReRAM channel, we propose an across-array coding strategy that assigns  a  codeword to multiple crossbar arrays. Since the coded bits at different arrays experience independent channels, the sneak-path rate within one codeword will be close to its mean value and hence the channel is closer to an i.i.d. channel.  Based on the across-array coding strategy, we further investigate the ReRAM channel capacity bounds.
 As the sneak path is dependent on the data message, it is difficult to derive the exact capacity of the ReRAM channel.
 By treating the sneak-path interference as the i.i.d. noise during decoding, the ReRAM channel over multiple memory arrays resembles a block-varying
channel that we will define in Section~\ref{sec:channel model}, whose status does not depend on the input data.
We then derive the capacity bounds of the block-varying channel, which can be regarded as an approximation of the ReRAM channel capacity.

\subsection{Across-Array Coding Strategy}
The proposed across-array coding strategy is illustrated in Fig.~\ref{fig:Coding}.
Consider the processing of the data vector $\fat{b}=(b_1, b_2,...,b_{NR})$, where $N$ is the code length and $R$ is the code rate. Here, $\fat{b}$ is encoded into codeword $\fat{x}=(x_1, x_2,...,x_N)$ which is assigned to $T$ memory arrays, where $N=sT$ for some integer $s$.  Thus, we split $\fat{x}$ into $T$ equal-length vectors, each of which is assigned to an independent memory array. Without loss of generality, we assign $\fat{x}^t=(x_1^t, x_2^t, ...., x_{N/T}^t)$ with $x_i^t=x_{(t-1)N/T+i}, i=1,..., N/T$, to the $t$-th memory array for $t=1,..., T$. Since each memory array is of size $m\times n$, $mnT/N$ codewords can be stored by these $T$ memory arrays, where $mnT/N$ is assumed to be an integer.  As the code rate is $R$, the storage efficiency is $R$ bits/cell.

Each codeword is decoded independently based on its readback signal. The codeword $\fat{x}=(\fat{x}^1,..., \fat{x}^T)$ is decoded based on its readback signal $\fat{y}=(\fat{y}^1,..., \fat{y}^T)$, where $\fat{y}^t$ is the readback signal of $\fat{x}^t$ from the $t$-th memory array, $t=1,..., T$, to obtain the estimated data $\hat{\fat{b}}$. Here we employ a suboptimal decoding scheme known as the TIN decoding \cite{Huang}. The TIN decoder ignores the correlation between the channel input and the sneak paths and treats the sneak-path interference as the i.i.d. noise.

\begin{figure*}[t]
\includegraphics[width=
6.0 in]{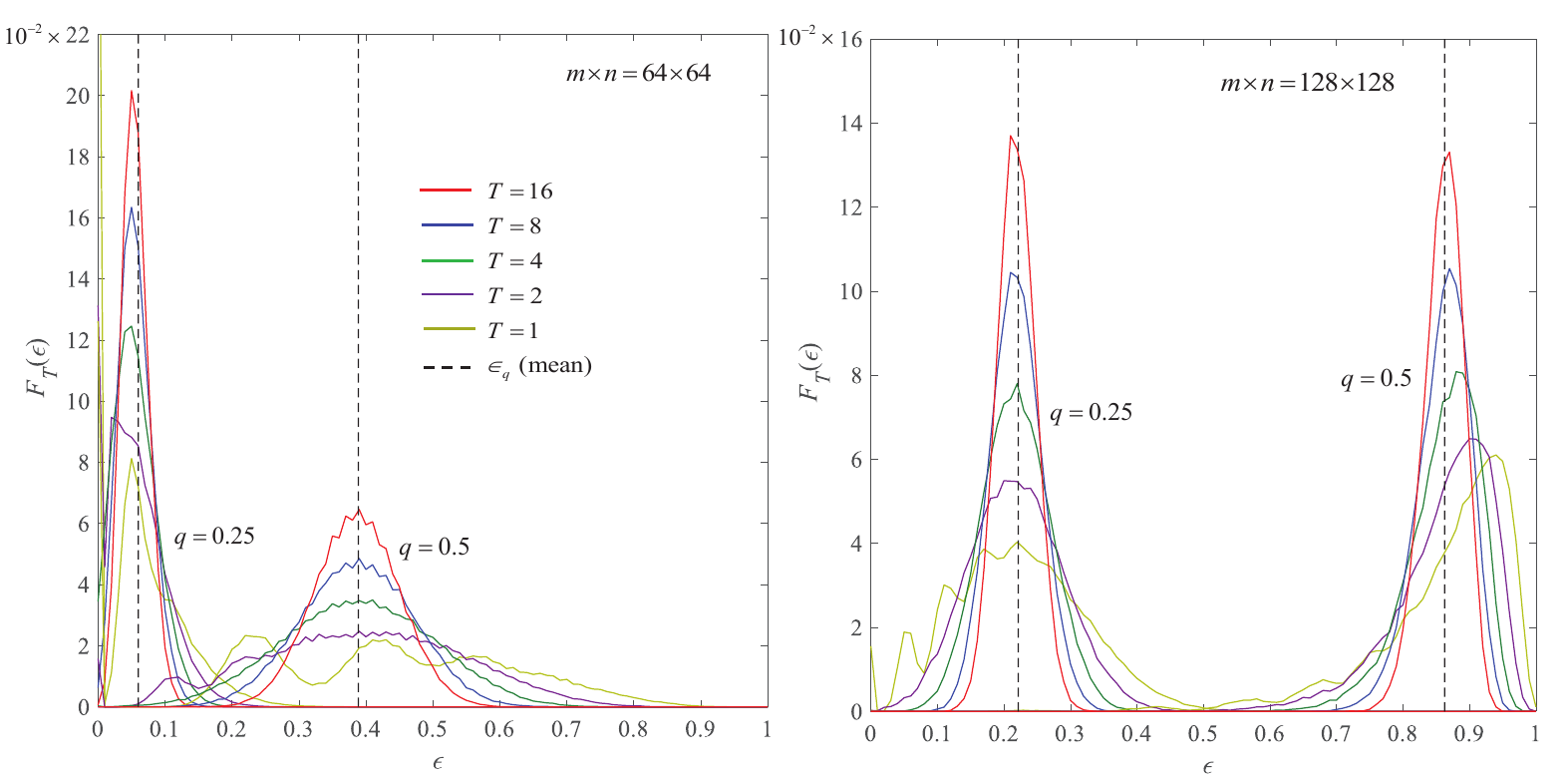}
\centering
\caption{PMFs of sneak-path rates over one codeword distributed in $T$ memory arrays with $T=1, 2, 4, 8, 16$. Code length and memory array size are  $m\times n= 64\times64$ (left) and $128\times128$ (right).} \label{fig:PDF64_128}
\end{figure*}

To illustrate the advantage of across-array coding strategy more explicitly, in Fig~\ref{fig:PDF64_128}, we evaluate the PMF of the sneak-path rate  over one codeword that is stored in $T$ memory arrays. In the figure, we employ the array sizes and the code lengths of $N=m\times n = 64\times64$ and $128\times128$. Since the coded bits are distributed in $T$ memory arrays, the sneak-path rate as well as its PMF are rewritten as $\sum_{t=1}^T\sum_{i=1}^{N/T}e_i^t/(mn(1-q))$ and $F^T_q(\epsilon)=\textrm{Pr}(\sum_{t=1}^T\sum_{i=1}^{N/T}e_i^t/(mn(1-q))=\epsilon)$, where $e_i^t$ is the sneak-path event indicator during the reading of the $i$-th bit that belongs to the $t$-th array.  For each case of $m\times n= 64\times64$ and $128\times128$ and the given input distribution with $q=0.25$ and $0.5$, as $T$ increases, the spread of the PMF of the sneak-path rate gets smaller and concentrates closer to the mean value $\epsilon_q$ and the channel becomes more stable. The reason is that since the codeword is assigned to $T$ independent memory arrays, the sneak-path rate is averaged over the $T$ arrays. Based on the law of large numbers, as $T\rightarrow\infty$, the sneak-path rate converges exactly to the mean value with probability 1, and therefore, we can design a code based on this mean value to guarantee error free decoding. Note that the across-array coding strategy does not change the channel correlation. It actually reduces the correlation of the sneak-path interference within one codeword since the readback signals for coded bits from different memory arrays are independent with each other. As all the coded bits are encoded from the same message data, this resembles a ``code diversity" strategy for block fading channels \cite{Raymond}.

The drawback of a joint coding across $T$ arrays is a potential increase of the read/write latency. Note that such additional latency is unavoidable for many sneak-path mitigating approaches in the literature. For example, the multistage reading technique presented in \cite{Vontobel} requires three readings and three writings to get a  better estimation of the sneak current; \cite{Ben} investigates the multiple-read detector and shows that it can achieve near-optimum performance with 10 reads. We remark that for our proposed across-array coding strategy, the additional read/write latency can be minimized if a parallel reading/writing circuit \cite{Zhou} is adopted across different crossbar arrays. Moreover, the additional latency incurred will become negligible if the two-dimensional crossbar arrays are stacked to form a 3D structure, which naturally enables the parallel reading/writing across different arrays \cite{Luo}.

\subsection{Channel Equivalence and Capacity Bounds}\label{sec:channel model}

We first define a block-varying $(\fat{\epsilon}, \sigma)$-channel and show how
the ReRAM channel capacity is related to that of the block varying
$(\fat{\epsilon}, \sigma)$-channel.

\begin{figure}[t]
\includegraphics[width=
3.3 in]{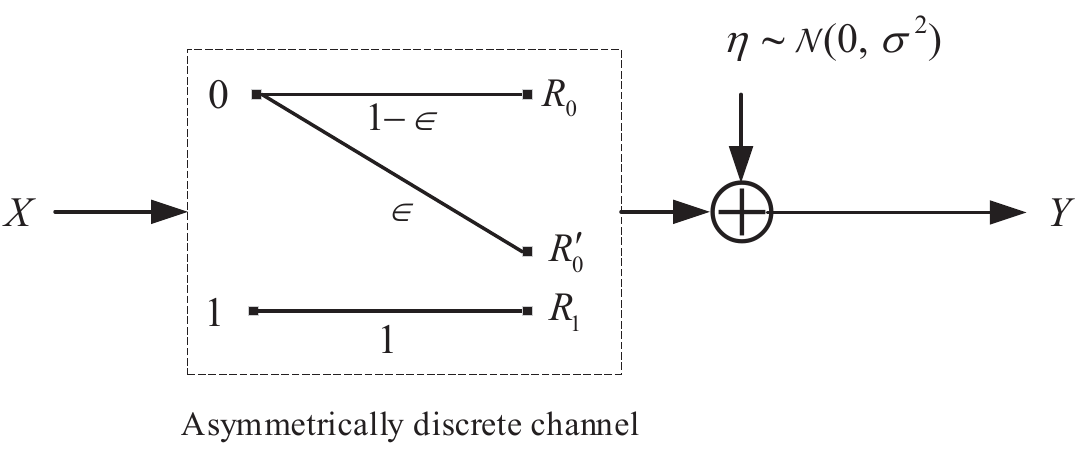}
\centering
\caption{$(\epsilon, \sigma)$-channel model with $R_0^\prime=\left(\frac{1}{R_0}+\frac{1}{R_s}\right)^{-1}$.} \label{fig:ESchannel}
\end{figure}

To begin, we first define an $(\epsilon, \sigma)$-channel.
As illustrated in Fig.~\ref{fig:ESchannel}, an $(\epsilon, \sigma)$-channel  is a concatenation of an i.i.d. asymmetrical discrete channel and an i.i.d. additive Gaussian channel, and therefore, it is also an i.i.d. channel without channel correlation. The asymmetrical discrete channel is with binary-input $X\in\{0, 1\}$ and ternary-output from $\{R_0, R_0^\prime, R_1\}$ with $R_0^\prime=\left(\frac{1}{R_0}+\frac{1}{R_s}\right)^{-1}$, and the transition property is described by Pr$(R_0|0)=1-\epsilon$, Pr$(R_0^\prime|0)=\epsilon$, and Pr$(R_1|1)=1$. The additive Gaussian channel is with noise distribution $\eta\sim\mathcal{N}(0, \sigma^2)$, whose output $Y\in \mathcal{R}$ serves as the output of the $(\epsilon, \sigma)$-channel.

 For given input distribution Pr$(X=0)=1-q$, Pr$(X=1)=q$, the $(\epsilon, \sigma)$-channel capacity can be derived as
\begin{eqnarray}
\!\!\!\!C_q(\epsilon, \sigma)&&\!\!\!\!\!\!\!\!=I(X;Y)\nonumber\\
&&\!\!\!\!\!\!\!\!=H(Y)-H(Y|X)\nonumber\\
&&\!\!\!\!\!\!\!\!=H(Y)-qH(Y|X=1)-(1-q)H(Y|X=0)\nonumber\\
&&\!\!\!\!\!\!\!\!=-\int_{-\infty}^{+\infty} p_Y(y)\log_2 p_Y(y)dy-q\log_2\sqrt{2\pi e\sigma^2}\nonumber\\
&&\!\!\!\!\!\!\!\! \ \ \ \ \ \ \ \ \ +(1-q)\int_{-\infty}^{+\infty} p_{Y|X=0}(y)\log_2 p_{Y|X=0}(y)dy\nonumber
\end{eqnarray}
where
\begin{eqnarray}
p_Y(y)=\!\!\!\!\!\!\!\!&&(1-q)\left(\epsilon \phi(y, R_0^\prime)+(1-\epsilon)\phi(y, R_0)\right)+q\phi(y, R_1)\nonumber\\
p_{Y|X=0}(y)=\!\!\!\!\!\!\!\!&&\epsilon \phi(y, R_0^\prime)+(1-\epsilon)\phi(y, R_0)\nonumber\\
\phi(y, m)=\!\!\!\!\!\!\!\!&&{1}/{(\sqrt{2\pi}\sigma)}e^{-\frac{(y-m)^2}{2\sigma^2}}.\nonumber
\end{eqnarray}

The $(0, \sigma)$- and $(1, \sigma)$-channels are asymmetrical binary-input AWGN channels, which are two special cases of an $(\epsilon, \sigma)$-channel. Obviously, the $(\epsilon, \sigma)$-channel capacity decreases as $\epsilon$ increases leading to $C_q(1, \sigma)<C_q(\epsilon, \sigma)<C_q(0, \sigma)$.

A $T$-block block-varying $(\fat{\epsilon}, \sigma)$-channel with parameters $\fat{\epsilon}=(\epsilon^1,\epsilon^2,...,\epsilon^T)$ varies from data block to data block, while within the $t$-th data block, the channel is a symbol-wise i.i.d. $(\epsilon^t, \sigma)$-channel, $t=1, 2,...,T$. The block-varying $(\fat{\epsilon}, \sigma)$-channel is a type of channel with block interference as proposed by \cite{Mceliece}.

The ReRAM channel over $T$ memory arrays resembles the $T$-block block-varying $(\fat{\epsilon}, \sigma)$-channel. In particular, the sneak-path rate of the ReRAM channel  varies from memory array to memory array resembles the block-varying property (the channel varies from block to block) of the  block-varying $(\fat{\epsilon}, \sigma)$-channel. The channel parameter $\epsilon^t=\textrm{Pr}(R_0^\prime|X=0)$ resembles the instantaneous sneak-path rate of the $t$-th memory array. The main difference is that in the ReRAM channel, the sneak-path interference is dependent on the input data and this data-dependency leads to the channel correlation, while the parameter $\fat{\epsilon}$ of block-varying $(\fat{\epsilon}, \sigma)$-channel is data-independent and the channel within each block is i.i.d. However, if we adopt TIN decoding where the decoder ignores this data-dependency and regards the sneak paths as the i.i.d. noise, an ReRAM channel is equivalent to a block-varying $(\fat{\epsilon}, \sigma)$-channel.  The memory block  length, denoted by $M$, of the  block-varying $(\fat{\epsilon}, \sigma)$-channel should be identical to the data array size of the ReRAM channel, i.e., $M=mn$. The channel parameters $\epsilon^t, t=1, 2, ..., T$ should be i.i.d. generated based on PMF of the sneak-path rate of the ReRAM channel. Therefore, the maximum achievable coding rate over the ReRAM channel under TIN decoding can be approximated by the block-varying $(\fat{\epsilon}, \sigma)$-channel capacity.

In preparation to give the capacity limit, we define an $(\epsilon, \sigma)$-channel code:
\begin{definition}\label{def:code}
An $(\epsilon, \sigma)$-channel code includes a sequences of codes with rate $C_q(\epsilon, \sigma)$ and different code lengths $n$, which achieve asymptotical error free decoding over the i.i.d. $(\epsilon, \sigma)$-channel as the code length approaches infinity.
\end{definition}

The existence of the $(\epsilon, \sigma)$-channel code ensemble is guaranteed by the conventional channel coding theorem.

 Consider a block-varying $(\fat{\epsilon}, \sigma)$-channel with memory block size $M$. Parameter $\epsilon^t$ has the PMF $F(\epsilon^t), t=1, 2, ..., T$, and mean value $\bar{\epsilon}=\sum_{\epsilon}\epsilon F(\epsilon)$, as defined in  (\ref{eq:SP-rate}). Since the channel varies from block to block, we consider joint $T$-block encoding and decoding. The code length is therefore $MT$. Let $R$ be the encoding rate, and we then have the following theorem:
\begin{theorem}\label{thm:thm}
For fixed input distribution of Bernoulli $(q)$, as $T\rightarrow \infty$, the maximum achievable code rate $R$ with joint $T$-block encoding and decoding  is bounded by
\begin{equation}
C_q(\bar{\epsilon}, \sigma)\leq R\leq \overline{C}_q(\sigma)
\end{equation}
where $\overline{C}_q(\sigma)=\sum_{\epsilon} F(\epsilon)C_q(\epsilon, \sigma)$.
\end{theorem}

\emph{Proof:}
We first show that for a fixed input distribution of Bernoulli $(q)$, $C_q(\bar{\epsilon}, \sigma)$ is achievable. Let $\fat{x}=(\fat{x}^1,..., \fat{x}^T)$ be the joint $T$-block codeword, where $t$-th block $\fat{x}^t=(x^t_1, ..., x^t_M)$ experiences an $(\epsilon^t, \sigma)$-channel, i.e., each symbol $x^t_j, j=1,..., M$, experiences an i.i.d. $(\epsilon^t, \sigma)$-channel.   We assume that the codeword is encoded in the way that the $i$-th bits, located at different data blocks, i.e., $(x_i^1, x_i^2, ..., x_i^T)$, belong to a codeword of a length-$T$ $(\bar{\epsilon}, \sigma)$-channel  code. In this way, the original length-$MT$ codeword $\fat{x}$ can be considered as a vector consisting $M$ length-$T$ $(\bar{\epsilon}, \sigma)$-channel codewords. This is possible because we can always split the uncoded data vector into $M$ equal-length sub-vectors, and encode each of them independently using an $(\bar{\epsilon}, \sigma)$-channel code. As the encoding rate of each $(\bar{\epsilon}, \sigma)$-channel code is $C_q(\bar{\epsilon}, \sigma)$, the overall code rate is $R=C_q(\bar{\epsilon}, \sigma)$.

During decoding, for each $i=1,..., M$, $(x_i^1, x_i^2, ..., x_i^T)$ is decoded based on its channel output $(y_i^1, y_i^2, ..., y_i^T)$, where $y_i^t$ is a channel observation of $x_i^t$. Since coded bit $x_i^t$ experiences an $(\epsilon^t, \sigma)$-channel, where $\epsilon^t, t=1, 2,..., T$, are i.i.d. generated based on the PMF of $F(\epsilon)$, the overall codeword experiences an $(\bar{\epsilon}, \sigma)$-channel where $\bar{\epsilon}=\sum_{\epsilon}\epsilon F(\epsilon)$. Since $(x_i^1, x_i^2, ..., x_i^T)$ is an $(\bar{\epsilon}, \sigma)$-channel codeword, the decoding error probability approaches 0 as $T\rightarrow\infty$ according to Definition~\ref{def:code}.

In \cite{Mceliece}, an upper bound of the block interference channel is proposed. That is, given the channel parameters $\fat{\epsilon}$, the channel becomes memoryless, leading to $R\leq\frac{1}{MT} I(\fat{x};\fat{y})\leq\frac{1}{MT} I(\fat{x};\fat{x}|\fat{\epsilon})\leq I(x_1^1;y_1^1|\epsilon^1) =\sum_{\epsilon}F(\epsilon) C_q(\epsilon, \sigma)=\overline{C}_q(\sigma)$.
\myQED

It was also shown in \cite{Mceliece} that  if the channel state is finite, i.e., $\epsilon$ is from a finite set, the upper bound is
tight when the block size  $M\rightarrow\infty$.

 \begin{figure*}[t]
\includegraphics[width=
6.2 in]{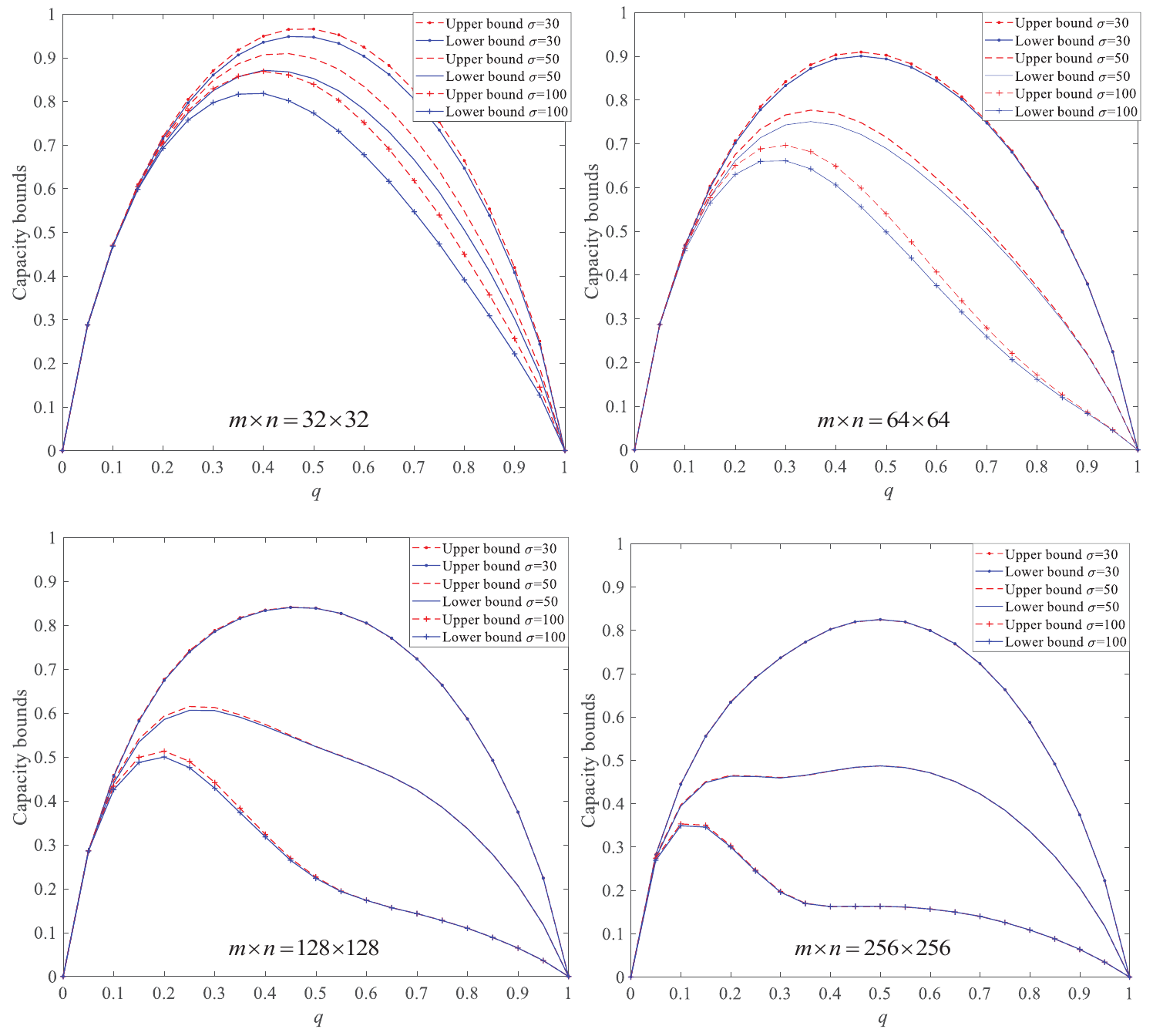}
\centering
\caption{Upper bound $\sum_\epsilon F_q(\epsilon)C_q(\epsilon, \sigma)$ and lower bound $C(\epsilon_q, \sigma)$ of the capacity as $T\rightarrow\infty$, with $R_1=100\ \Omega, R_0=1000\  \Omega, R_s=250\ \Omega$, and $p_f=10^{-3}$.} \label{fig:capacity}
\end{figure*}

Based on our channel equivalence, $\max_qC_q(\epsilon_q, \sigma)$ is an approximate lower bound of the ReRAM channel capacity with  TIN decoding, and $\max_q\sum_\epsilon F_q(\epsilon)C_q(\epsilon, \sigma)$ is an upper bound. Since we have the explicit formula (\ref{eq:SP-rate-cal}) for $\epsilon_q$, the lower bound is much easier to be calculated than the upper bound, which requires $F_q(\epsilon)$.
Fortunately, we can show that when the array size $N$ is large, the two bounds are numerically very close with each other, and hence the lower bound $\max_qC(\epsilon_q, \sigma)$ can be a good approximation of the ReRAM channel capacity. Fig.~\ref{fig:capacity} illustrates the capacity upper bound $\sum_\epsilon F_q(\epsilon)C_q(\epsilon, \sigma)$ and lower bound $C(\epsilon_q, \sigma)$ as functions of $q$, for different memory array sizes $m\times n= 32\times32, 64\times64, 128\times128, 256\times 256$ and different noise values of $\sigma=30, 50, 100$. The resistance parameters are fixed with $R_1=100\ \Omega, R_0=1000\  \Omega$, and $R_s=250\ \Omega$. Observe that when the memory array size $N$ is large, the two bounds are very close with each other.

Fig.~\ref{fig:capacity}  also indicates that the ReRAM channel capacity bounds decrease as the data size increases due to the increase of the sneak-path rate, i.e., the larger the data array size the lower the average storage efficiency for each cell, and vice versa. For a very low noise level of $\sigma=30$, the capacity bounds are maximized at about $q=0.5$, and for $\sigma=50, 100$, they are typically maximized when $q<0.5$. The optimal value of $q$ that maximizes the channel capacity bounds decreases as noise level increases. This is because noise amplifies the detrimental effect of sneak paths, while reducing $q$ effectively reduces the sneak-path rate.

\section{Design of the Coding Scheme}\label{sec:coding}
In this section, we present the design of a coding scheme for the ReRAM channel. We also propose a real-time maximum likelihood channel estimation, based on which the message-passing decoding is performed. We utilize the state-of-the-art sparse-graph code and message-passing decoding theories to design the coding scheme, which is essentially an $(\epsilon_q, \sigma)$-channel code design. A major difference between the $(\epsilon_q, \sigma)$-channel code and the classical ECC is that since the former works over an asymmetrical channel, the coded data should follow the desired distribution to approach the channel capacity  (Fig.~\ref{fig:capacity}). We propose a coding scheme, which is a serial concatenation between a classical ECC and a data shaper that shapes the desired data distribution. Bit error rate (BER) simulations and performance comparisons are also presented in this section for our proposed coding scheme.
\begin{figure*}[t]
\includegraphics[width=
4.8 in]{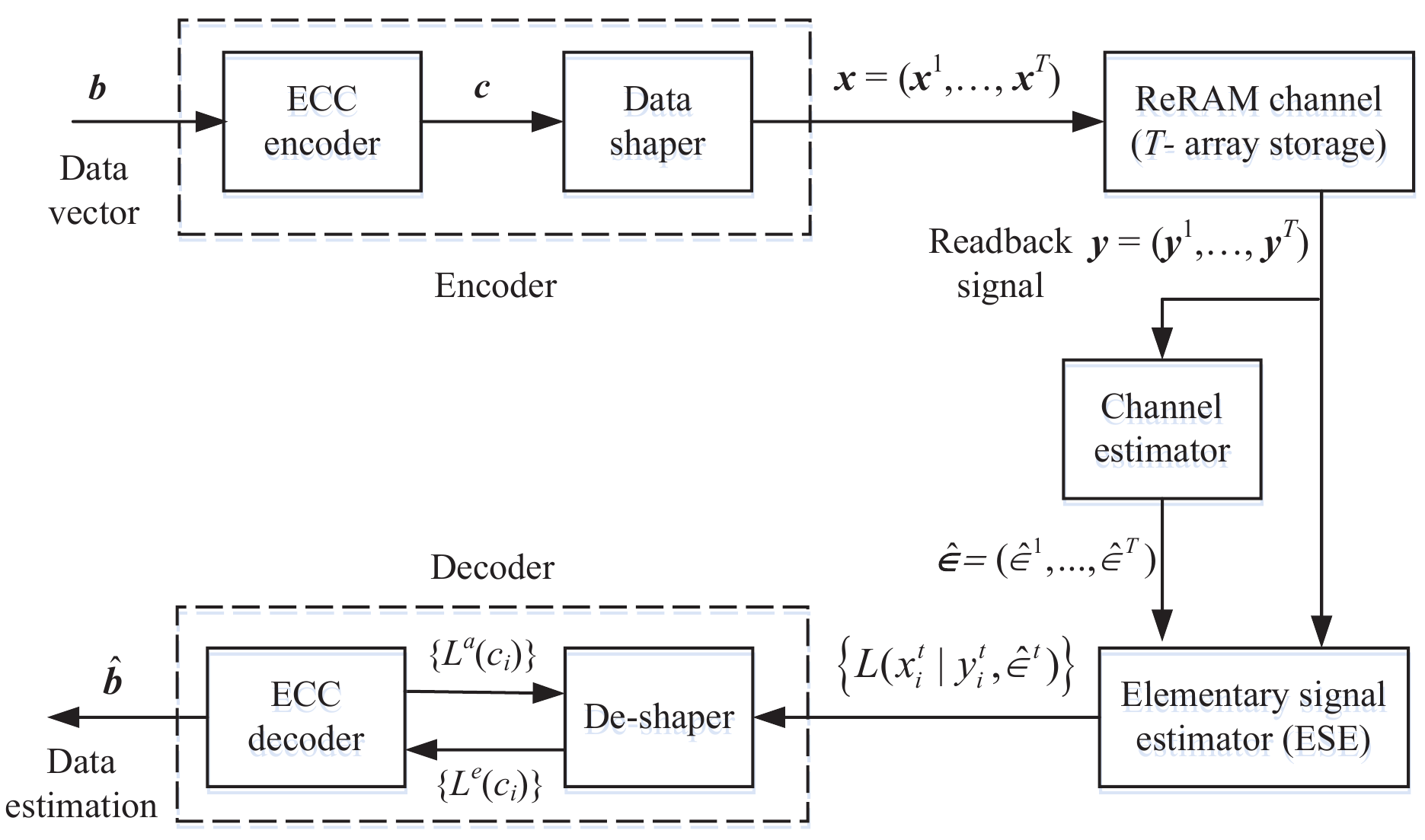}
\centering
\caption{Proposed coding scheme for the ReRAM channel.} \label{fig:Code_shape}
\end{figure*}
\subsection{Coding and Decoding Model}
A system model for the proposed coding scheme is illustrated in Fig.~\ref{fig:Code_shape}. The encoder includes an ECC encoder and a data shaper. The ECC encoder encodes data vector $\fat{b}=(b_1, b_2,...,b_{NR})$ into codeword $\fat{c}=(c_1, c_2,...,c_N)$ whose entries are uniformly distributed on $\{0, 1\}$. The data shaper reforms the data distribution into Bernoulli $(q)$. Its output is $\fat{x}=(x_1, x_2,...,x_N)$. Here the data shaper in our system has rate-1, and therefore, the overall code rate is still $R$.

The decoder involves a real-time channel estimator, elementary signal estimator (ESE), a de-shaper, and an ECC decoder. Since the decoding is actually a block-varying $(\fat{\epsilon}, \sigma)$-channel decoding, the channel estimator first estimates the channel parameters $\fat{\epsilon}=(\epsilon^1,...,\epsilon^T)$ over the $T$ memory arrays based on readback signal $\fat{y}=(\fat{y}^1,..., \fat{y}^T)$. Based on $\hat{\fat{\epsilon}}$ and $\fat{y}$, the ESE calculates a soft estimation $\{L(x_i^t|y_i^t, \hat{\epsilon}^t)\}$, i.e., the log-likelihood ratio (LLR), for each coded bit $x_i^t$ that is used as the decoder input. The decoder consists of a de-shaper and an ECC decoder, both of which use soft-in soft-out (SISO) processings and perform iteratively to improve the decoding reliability. The corresponding decoding is standard message-passing decoding. Specifically, the de-shaper calculates soft LLR $\{L^e(c_i)\}$ for each ECC coded bit, based on which an ECC decoding refines the estimation and feds back an updated LLR $\{L^a(c_i)\}$  to the de-shaper for the next round of decoding iterations. After a fixed maximum number of iterations, a hard decision is performed  at the ECC decoder to produce data estimation $\hat{\fat{b}}$.

\begin{figure}[t]
\includegraphics[width=
3.5 in]{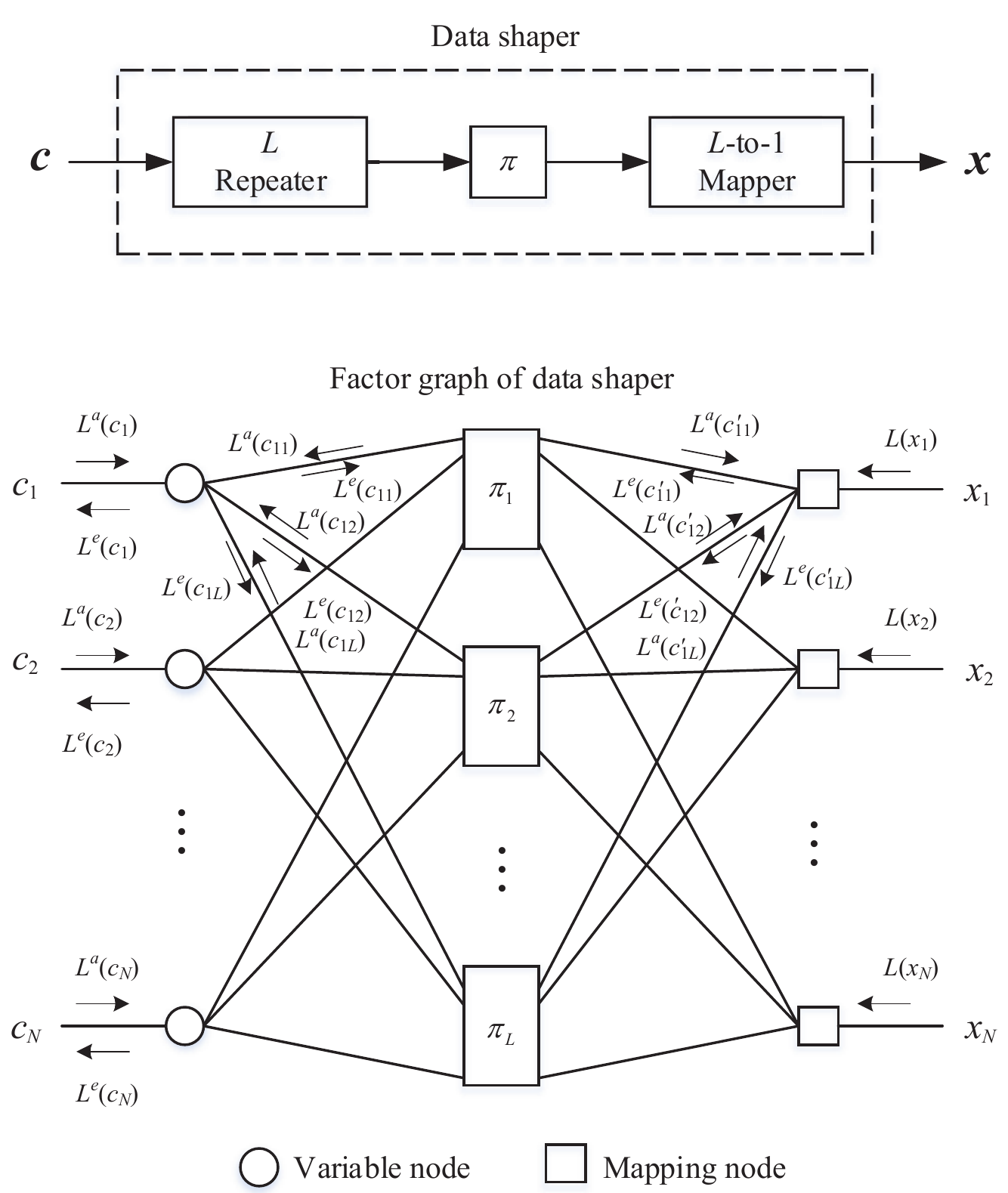}
\centering
\caption{Data shaper and its factor graph illustration.} \label{fig:Shaper}
\end{figure}

\subsection{Data Shaper}\label{sec:shaper}
The data shaper consists of a length-$L$ repeater, a length-$NL$ bit interleaver $\pi$, and an $L$-to-1 mapper (Fig.~\ref{fig:Shaper}). The repeater duplicates each ECC coded bit $L$ times. The bit interleaver permutes the repeater output to relocate the bits. The $L$-to-1 mapper maps every $L$ bits to one bit, i.e., $\mathcal{M}: \{0, 1\}^L\rightarrow \{0, 1\}$. Therefore, the data shaper's overall rate is 1. For the $L$-to-1 mapper, since there are $2^L$ patterns for the mapping inputs, by mapping $i$ of them to 1 and $2^L-i$ of them to 0, we obtain the data output with a distribution of Bernoulli $(\frac{i}{2^L})$. By choosing $i=1, 2,..., 2^L-1$, we achieve data distributions of Bernoulli $(q)$ with $q=\frac{1}{2^L}, \frac{2}{2^L}, ...,  \frac{2^L-1}{2^L}$.

\begin{figure}[t]
\includegraphics[width=
1.3 in]{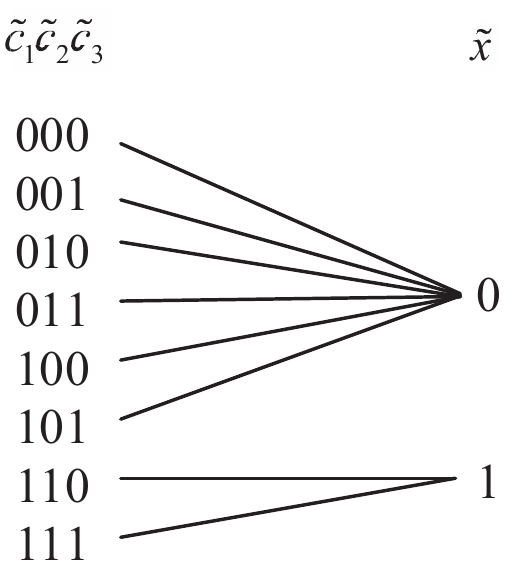}
\centering
\caption{3-to-1 mapping for output data distribution Pr$(\tilde{x}=0)=\frac{3}{4}$, Pr$(\tilde{x}=1)=\frac{1}{4}$.} \label{fig:map3}
\end{figure}

The interleaver inside the data shaper is crucial in our scheme. Rather than adopting random interleaving, we propose a structured interleaving scheme, as shown in the data shaper's factor graph in Fig.~\ref{fig:Shaper}. The interleaver $\pi$ consists of $L$ sub-interleavers $\pi_i, i=1,..., L$, each of which can be random. The data-shaping process can be described using a factor graph. Each variable node is associated with an ECC coded bit, where the $i$-th variable node is associated with $c_i$. There are $L$ edges from a variable node to the interleavers corresponding to the $L$ repetitions of the ECC coded bit. Each mapping node has $L$ edges from the interleavers corresponding to the $L$ mapping inputs. The $i$-th mapping node is associated with the mapping output $x_i$. By using our structured interleaver, the $i$-th repetitions of the ECC coded bits enter a sub-interleaver $\pi_i$ whose outputs are used as the $i$-th inputs of the mapping nodes. By doing so, each ECC coded bit has exactly one repetition that occupies the $i$-th input of a mapping node for $i=1, ..., L$.

The advantage of employing this structured interleaver can be explained using an example. Consider a data shaper with an $(L=3)$-repeater and a 3-to-1 mapper $\mathcal{M}(\tilde{c}_1,\tilde{c}_2,\tilde{c}_3)=\tilde{x}$ (Fig.~\ref{fig:map3}). There are eight patterns for three binary inputs $\tilde{c}_1,\tilde{c}_2,\tilde{c}_3$, where only two of them, 110 and 111, are mapped to 1, and the other six are mapped to 0. If  $\tilde{c}_1,\tilde{c}_2,\tilde{c}_3$ are i.i.d. with Pr$(\tilde{c}_i=0)$=Pr$(\tilde{c}_i=1)=\frac{1}{2}, i=1, 2, 3$, the  mapping can realize output data distribution with Pr$(\tilde{x}=0)=\frac{3}{4}$, Pr$(\tilde{x}=1)=\frac{1}{4}$. Next we address the de-mapping. Mapping output $\tilde{x}$ actually contains a different quantity of information about the three input bits $\tilde{c}_1,\tilde{c}_2,\tilde{c}_3$. By formulating the mapping rule as $\tilde{x}=\mathcal{M}(\tilde{c}_1,\tilde{c}_2,\tilde{c}_3)=\tilde{c}_1\cdot\tilde{c}_2$, where $\cdot$ is a multiply operation, we evaluate the mutual information between $\tilde{x}$ and each input bit as $I(\tilde{x};\tilde{c}_1)=I(\tilde{x};\tilde{c}_2)=\frac{3}{4}\log_2\frac{4}{3}$ and  $I(\tilde{x};\tilde{c}_3)=0$.
Therefore, if the de-mapper is sufficiently near-optimal, during de-mapping we can obtain information about the first and second bits $\tilde{c}_1, \tilde{c}_2$. Unfortunately, we cannot get any information about the third bit $\tilde{c}_3$ since $\tilde{x}$ does not contain any information about $\tilde{c}_3$. In other words, one-third of the bits are erased after de-mapping. If random interleaving is employed, with probability $\frac{1}{27}$, all the three repetitions of an ECC coded bit will be assigned as the third input of a mapping node and erased. In other words, with probability $\frac{1}{27}$, an ECC coded bit will be erased after de-shaping, which leads to a poor ECC decoding performance.
Our structured interleaving guarantees that all the ECC coded bits can obtain a positive and statistically equal quantity of information from the de-shaper to benefit the ECC decoding.

\subsection{Channel Estimation and ESE}
We propose a maximum likelihood channel estimation to obtain parameters $\fat{\epsilon}=(\epsilon^1,...,\epsilon^T)$.
Since the decoder assumes the channel created by the $t$-th memory array as an i.i.d. $(\epsilon^t, \sigma)$-channel, with the channel observation of $\fat{y}^t=(y^t_1,...,y^t_{N/T})$, the log-likelihood function of $\epsilon^t$ is written as:
\begin{eqnarray}
\Lambda(\epsilon^t;\fat{y}^t)
&&\!\!\!\!\!\!\!\!\!\!=\log \prod_{i=1}^{N/T}\textrm{Pr}\left(y^t_i|\epsilon^t\right)\\
&&\!\!\!\!\!\!\!\!\!\!=\sum_{i=1}^{N/T}\log\left[ (1-q)\left(\epsilon^t \phi(y^t_i, R_0^\prime)+(1-\epsilon^t)\phi(y^t_i, R_0)\right)\right.\nonumber\\
&&\!\!\!\!\!\!\!\!\!\!\ \ \ \ \ \ \left.+q\phi(y^t_i, R_1)\right].  \label{eq:logsum}
\end{eqnarray}
Next we consider the approximation for (\ref{eq:logsum}). Let $\bar{y}^t_i=\arg\min_{x\in\{R_0, R_0^\prime, R_1\}}|y^t_i-x|$ be the hard decision value of $y^t_i$.  Since each term of (\ref{eq:logsum}) is in the form of $\log\sum_{x\in\{R_0, R_0^\prime, R_1\}}p_x\phi(y^t_i, x)$,  when the channel noise level is low, it is dominated by the term of $x=\bar{y}^t_i$. We thereby apply
\begin{eqnarray}
\log\sum_{x\in\{R_0, R_0^\prime, R_1\}}p_x\phi(y^t_i, x)&&\!\!\!\!\!\!\!\!\!\!\approx \log \left(p_{\bar{y}^t_i}\phi(y^t_i, \bar{y}^t_i)\right)\nonumber\\
&&\!\!\!\!\!\!\!\!\!\!=\log p_{\bar{y}^t_i}-\frac{(y^t_i-\bar{y}^t_i)^2}{2\sigma^2}-\frac{1}{2}\log(2\pi\sigma^2),\nonumber
\end{eqnarray}
and hence have the following approximation:
\begin{eqnarray}
\Lambda(\epsilon^t;\fat{y}^t)\approx n_{R_0^\prime}\log\epsilon^t+n_{R_0}\log(1-\epsilon^t)-\sum_{i=1}^{N/T}\frac{(y^t_i-\bar{y}^t_i)^2}{2\sigma^2}\!+\!c\nonumber
\end{eqnarray}
where $n_x\overset{\Delta}{=}\sum_{i=1}^{N/T}1\{\bar{y}^t_i=x\}$ is the total number of $y^t_i$ whose hard decision is $x$ and $c$ is a constant term independent of $\epsilon^t$. By maximizing $\Lambda(\epsilon^t;\fat{y}^t)$ we obtain the estimation of $\epsilon^t$:
\begin{eqnarray}
\hat{\epsilon}^t=\arg\max_{\epsilon^t}\Lambda(\epsilon^t;\fat{y}^t)\approx \frac{n_{R_0^\prime}}{n_{R_0^\prime}+n_{R_0}}.\label{eq:channel_est}
\end{eqnarray}

Next, the ESE calculates the LLR for each coded bit of $(\fat{x}^1,..., \fat{x}^T)$ based on the readback signal $(\fat{y}^1,..., \fat{y}^T)$ and the estimated channel parameters:
     \begin{eqnarray}
   L(x_i^t|y_i^t, \hat{\epsilon}^t) &&\!\!\!\!\!\!\!\!\!\!=\log\frac{\textrm{Pr}(y_i^t|x_i^t=0, \hat{\epsilon}^t)}{\textrm{Pr}(y_i^t|x_i^t=1, \hat{\epsilon}^t)}\\
    &&\!\!\!\!\!\!\!\!\!\!=\log\frac{\hat{\epsilon}^t \phi(y_i^t, R_0^\prime)+(1-\hat{\epsilon}^t)\phi(y_i^t, R_0)}{\phi(y_i^t, R_1, \sigma^2)}
   \end{eqnarray}
   for $i=1,..., N/T, t=1,...,T$.
   Note that the implementation of ESE (\ref{eq:ESE})  ignores the correlation between cells in a memory array and regards the sneak-path interference as the i.i.d. noise.
A more sophisticated decoding scheme can be developed by utilizing the cell correlation and performing joint data and sneak path detection. It is left as our future work.

   \begin{figure}[t]
\includegraphics[width=
3.5 in]{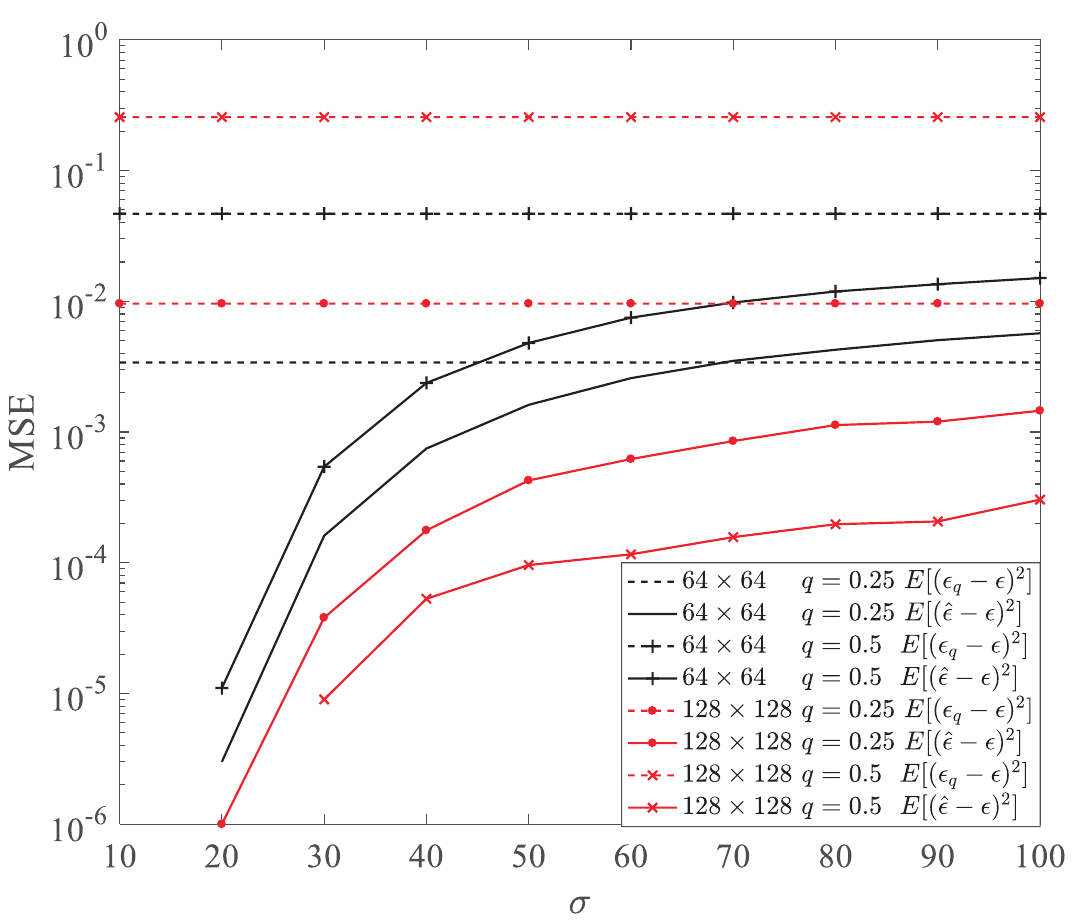}
\centering
\caption{MSE: $E[(\hat{\epsilon}-\epsilon)^2]$ of our proposed channel estimation (solid lines) and MSE: $E[(\epsilon_q-\epsilon)^2]$ of the average channel parameter (dashed lines). } \label{fig:Channel_mse}
\end{figure}

To demonstrate the accuracy of the proposed channel estimation, in Fig.~\ref{fig:Channel_mse}, we illustrate the mean squared error (MSE) between the estimated and the actual sneak-path rates. We obtain the MSE: $E[(\hat{\epsilon}-\epsilon)^2]$ by simulation for $T=1$ and memory array sizes $m\times n=64\times64, 128\times128$, where $\epsilon$ is the actual sneak-path occurrence rate, and $\hat{\epsilon}$ is the estimated value obtained by (\ref{eq:channel_est}). In general, the MSE is below $10^{-2}$ and decreases as the channel noise level decreases. For comparison, we also illustrate the MSE: $E[(\epsilon_q-\epsilon)^2]$ between the average and the actual sneak-path rates, where the average sneak-path rate is employed by the decoder when channel estimation is unavailable. Our proposed channel estimation is much more accurate to predict the channel than the average channel parameter, especially for large array size.

\subsection{De-Shaper}

The SISO de-shaper can also be realized by message-passing processing over the factor graph shown in Fig.~\ref{fig:Shaper}. During the de-shaping, each node performs as a local processor and the edges pass LLR messages. The message passing on the edges is bi-directional. The overall processing is performed iteratively. In each iteration, each node in the factor graph acts once. A mapping node performs de-mapping processing based on the $L$ priori LLRs from its neighboring variable nodes and the LLR from ESE and outputs an extrinsic LLR for each mapping input. The extrinsic LLR is used as an a priori LLR for variable node processings. A variable node combines the $L$  priori LLRs from its neighboring mapping nodes and feeds back an extrinsic LLR to each of  its neighboring mapping nodes. After a certain number of iterations the variable nodes output a more reliable LLR for each ECC coded bit as the de-shaper output.

\subsubsection{Mapping Node Processing}
A mapping node represents a mapping constraint, i.e., the $L$ edges on the left side link to the $L$ variable nodes that are the $L$ mapping input, and the edge on the right side links to a mapping output. Therefore, the $i$-th mapping node represents mapping constraint $\mathcal{M}(c_{i,1}^\prime, c_{i,2}^\prime, \cdots, c_{i,L}^\prime)=x_i$, where $c_{i,1}^\prime, c_{i,2}^\prime, \cdots, c_{i,L}^\prime$ are the mapping inputs and $x_i$ is the mapping output. Thus, the edges from the left of a mapping node should pass the LLRs for $c_{i,1}^\prime, c_{i,2}^\prime, \cdots, c_{i,L}^\prime$ and the edge from the right side should pass LLR for $x_i$.

 Let $L(x_i)=\log\frac{\textrm{Pr}(y_i|x_i=0)}{\textrm{Pr}(y_i|x_i=1)}$ be the LLR about $x_i, i=1,...,N$, obtained from the ESE.
   Let $L^a(c_{i,j}^\prime)=\log\frac{\textrm{Pr}(c_{i,j}^\prime=0)}{\textrm{Pr}(c_{i,j}^\prime=1)}$ be an a priori LLR of $c_{i,j}^\prime$ from a variable node.
 The $i$-th mapping node calculates an extrinsic LLR for $c_{i,k}^\prime, k=1,2,...,L$, given by
 \begin{eqnarray}
 L^e(c_{i,k}^\prime)&&\!\!\!\!\!\!\!\!=\log\frac{\textrm{Pr}(y_i|c_{i,k}^\prime=0)}{\textrm{Pr}(y_i|c_{i,k}^\prime=1)}\nonumber\\
  &&\!\!\!\!\!\!\!\!=\log\frac{\sum_{c_{i,j}^\prime, j\neq k}\textrm{Pr}(y_i|c_{i,k}^\prime=0, c_{i,j}^\prime, j\neq k)\prod_{j\neq k }\textrm{Pr}(c_{i,j}^\prime)}{\sum_{c_{i,j}^\prime, j\neq k}\textrm{Pr}(y_i|c_{i,k}^\prime=1, c_{i,j}^\prime, j\neq k)\prod_{j\neq k }\textrm{Pr}(c_{i,j}^\prime)}\nonumber\\
 &&\!\!\!\!\!\!\!\!=\log\frac{\sum_{c_{i,1}^\prime, c_{i,2}^\prime, \cdots, c_{i,L}^\prime}(1-c_{i,k}^\prime)\textrm{Pr}(y_i|x_i)\prod_{j\neq k }\textrm{Pr}(c_{i,j}^\prime)}{\sum_{c_{i,1}^\prime, c_{i,2}^\prime, \cdots, c_{i,L}^\prime}c_{i,k}^\prime\textrm{Pr}(y_i|x_i)\prod_{j\neq k }\textrm{Pr}(c_{i,j}^\prime)}\nonumber\\
  &&\!\!\!\!\!\!\!\!=\log\frac{\sum_{c_{i,1}^\prime, c_{i,2}^\prime, \cdots, c_{i,L}^\prime}(1-c_{i,k}^\prime)e^{(1-x_i)L(x_i)+\sum_{j\neq k}(1-c_{i,j}^\prime)L^a(c_{i,j}^\prime)}}{\sum_{c_{i,1}^\prime, c_{i,2}^\prime, \cdots, c_{i,L}^\prime}c_{i,k}^\prime e^{(1-x_i)L(x_i)+\sum_{j\neq k}(1-c_{i,j}^\prime)L^a(c_{i,j}^\prime)}}\nonumber
 \end{eqnarray}
 where $x_i=\mathcal{M}(c_{i,1}^\prime, c_{i,2}^\prime, \cdots, c_{i,L}^\prime)$.

   \subsubsection{Variable Node Processing}
   Since a variable node is associated with an ECC coded bit, i.e., the $i$-th node is associated with $c_i$, the edges connected to it should pass LLR messages for the same bit, i.e., $c_{i,1}=c_{i,2}=\cdots=c_{i,L}=c_i$, where $c_{i,j}, j=1,..., L$, are $L$ repetitions of $c_i$.

 Consider the processing at the $i$-th variable node. Let $L^a(c_i)$ be the priori LLR about $c_i$ from the ECC decoder and $L^a(c_{i,j})$ be the priori LLR about $c_{i,j}, j=1,...,L$, from the neighboring mapping nodes. Since  $c_{i,1}=c_{i,2}= \cdots = c_{i,L}=c_i$,
   the variable node calculates an extrinsic LLR for each $c_{i,k}$, given by
     \begin{eqnarray}
   L^e(c_{i,k}) =L^a(c_i)+\sum_{j\neq k}L^a(c_{i,k}).
   \end{eqnarray}

After a certain number of processing iterations, the variable node outputs an extrinsic LLR about $c_i$ to the ECC decoder
     \begin{eqnarray}
   L^e(c_i) =\sum_{k=1}^LL^a(c_{i,k}).
   \end{eqnarray}

\subsection{ECC Optimization and BER Simulations}
 In this section, we present the
ECC optimization and BER simulation results for ReRAM systems. With our proposed ECC, we achieve a high storage efficiency with a gap of less than 0.1 bit/cell from the ReRAM channel capacity.

\begin{figure}[t]
\includegraphics[width=
3.3 in]{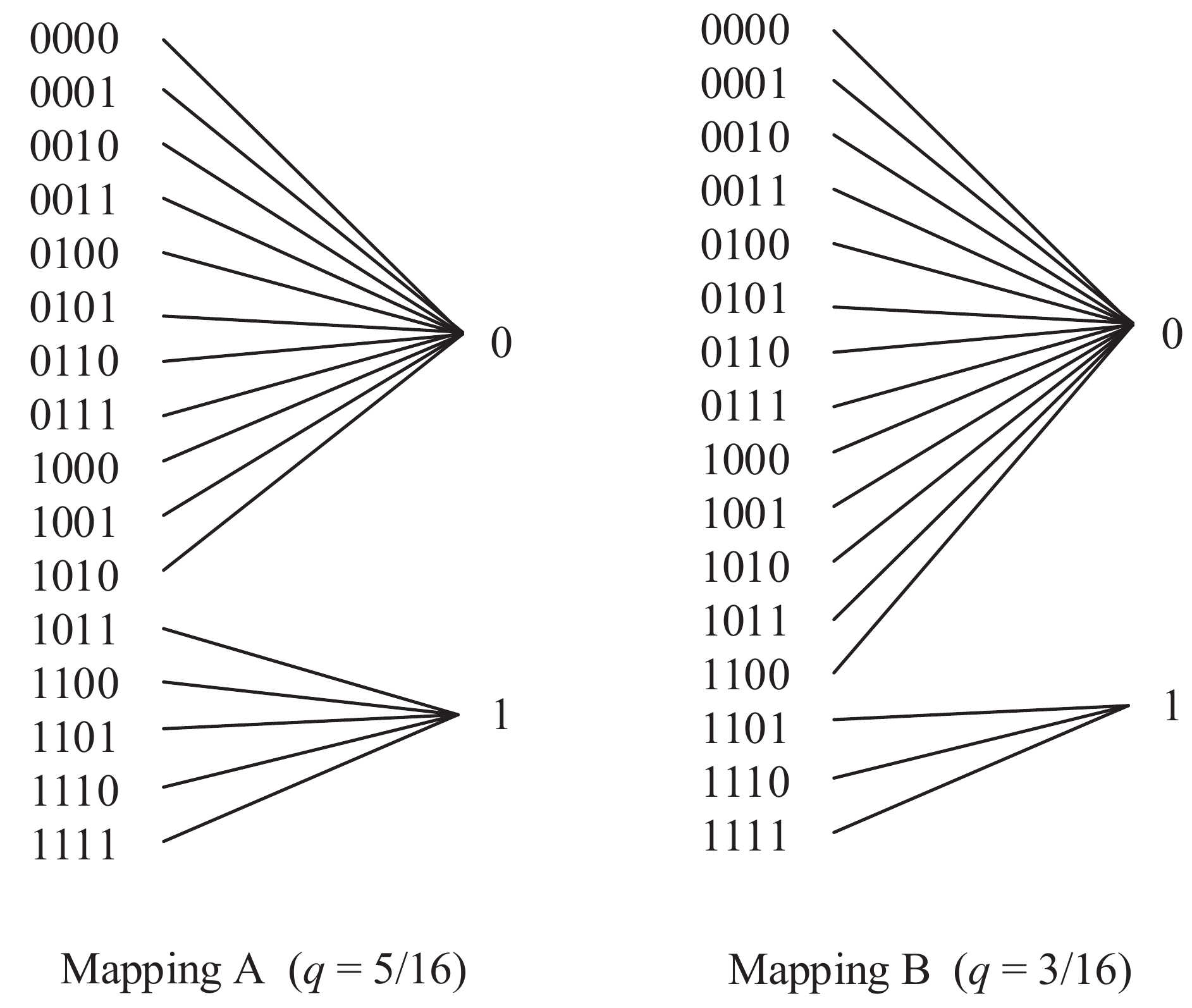}
\centering
\caption{Mappings A and B that achieved data distributions with $q=5/16$ and $q=3/16$.} \label{fig:Mapping4}
\end{figure}

\begin{table}[t]
\caption{Code parameters for $m\times n=64\times64$ and $128\times128$ ReRAM systems. Parameters of IRA code involves its variable node degree distribution: $\{\lambda_i\}$ and combiner factor: $a$ \cite{RA,IRA}.}
\label{tab:code}
\begin{center}
\begin{tabular}{|c|c|c|}
\hline\hline
Array size&\multirow{2}{*}{$64\times64$} &\multirow{2}{*}{$128\times128$}\\
$m\times n$ & & \\
\hline
\multirow{3}{*}{Data shaper}& $L=4$ & $L=4$\\
& Mapping A & Mapping B\\
& $q=5/16$ & $q=3/16$\\
\hline
\multirow{3}{*}{IRA code (ECC)} & $\lambda_3=0.567736$&$\lambda_3=0.501564$\\
& $\lambda_{50}=0.432264$&$\lambda_{50}=0.498436$\\
&Combiner: $a=6$&Combiner: $a=4$\\
\hline
Code rate &$R=0.542824$&$R=0.414735$\\
\hline
$\max_q(C_q(\epsilon_q, \sigma=100))$ &0.660 &0.494\\
\hline\hline
\end{tabular}
\end{center}
\end{table}

\begin{figure}[t]
\includegraphics[width=
3.5 in]{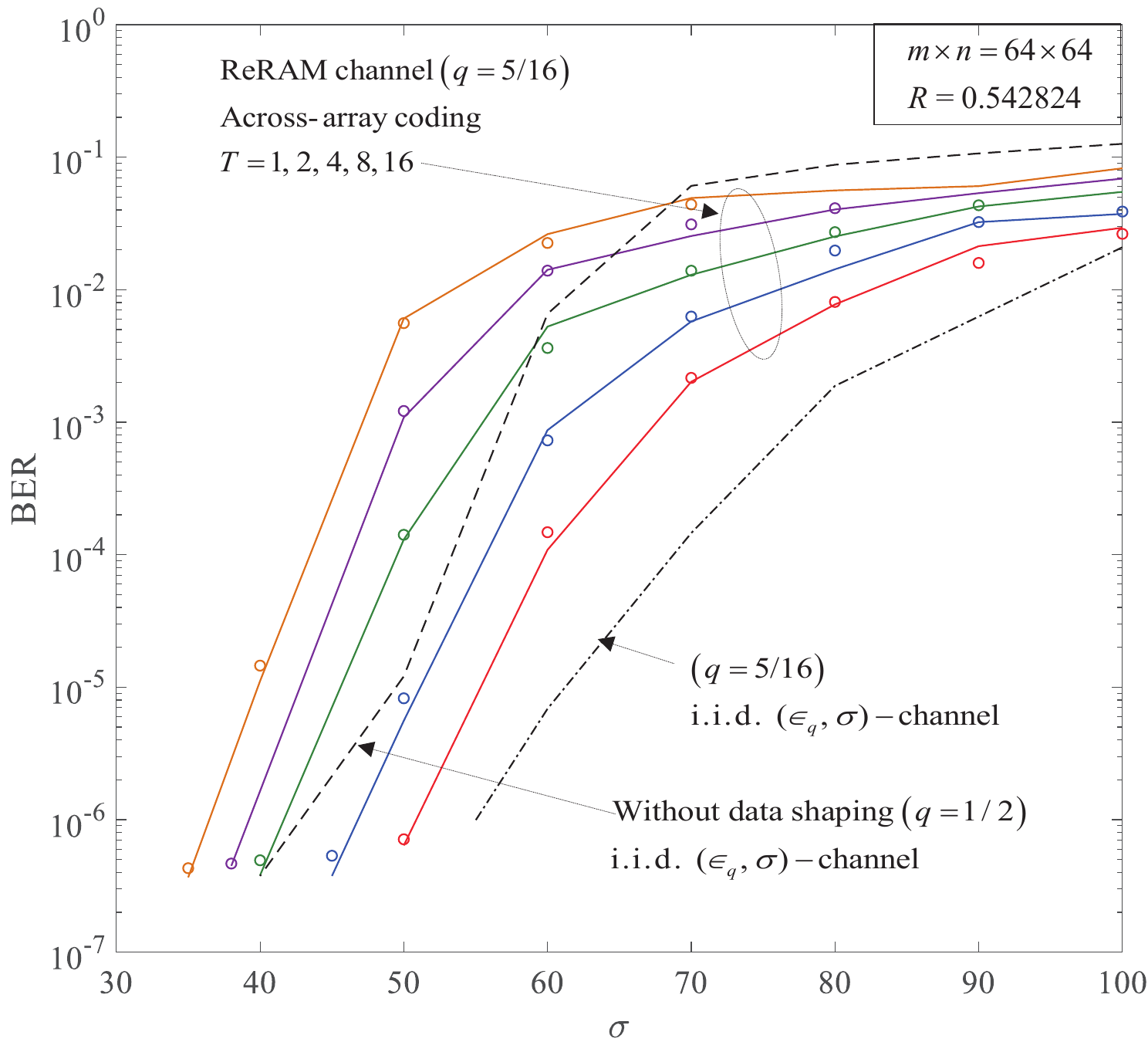}
\centering
\caption{BERs for IRA-coded ReRAM channel (solid line) and IRA-coded block-varying $(\fat{\epsilon}, \sigma)$-channel (labeled by $\circ$) with across-array coding strategy where channel parameters are set as $m\times n=64\times64$, $R_1=100\ \Omega, R_0=1000\  \Omega, R_s=250\ \Omega$,  and $p_f=10^{-3}$.} \label{fig:BER64}
\end{figure}

\begin{figure}[t]
\includegraphics[width=
3.5 in]{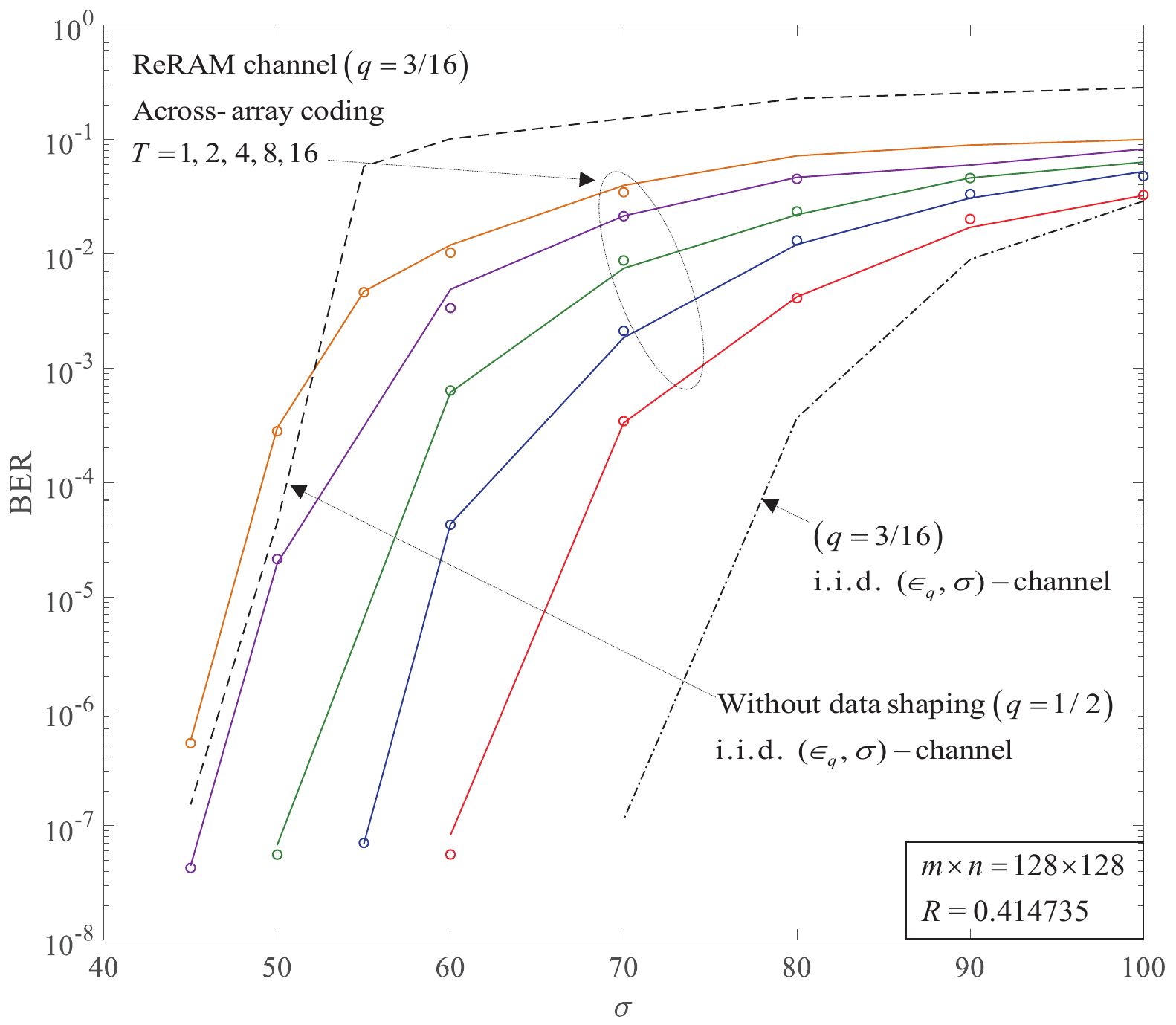}
\centering
\caption{BERs for IRA-coded ReRAM channel (solid line) and IRA-coded block-varying $(\fat{\epsilon}, \sigma)$-channel (labeled by $\circ$) with across-array coding strategy where channel parameters are set as $m\times n=64\times64$, $R_1=100\ \Omega, R_0=1000\  \Omega, R_s=250\ \Omega$,  and $p_f=10^{-3}$.} \label{fig:BER128}
\end{figure}

\begin{figure}[t]
\includegraphics[width=
3.5 in]{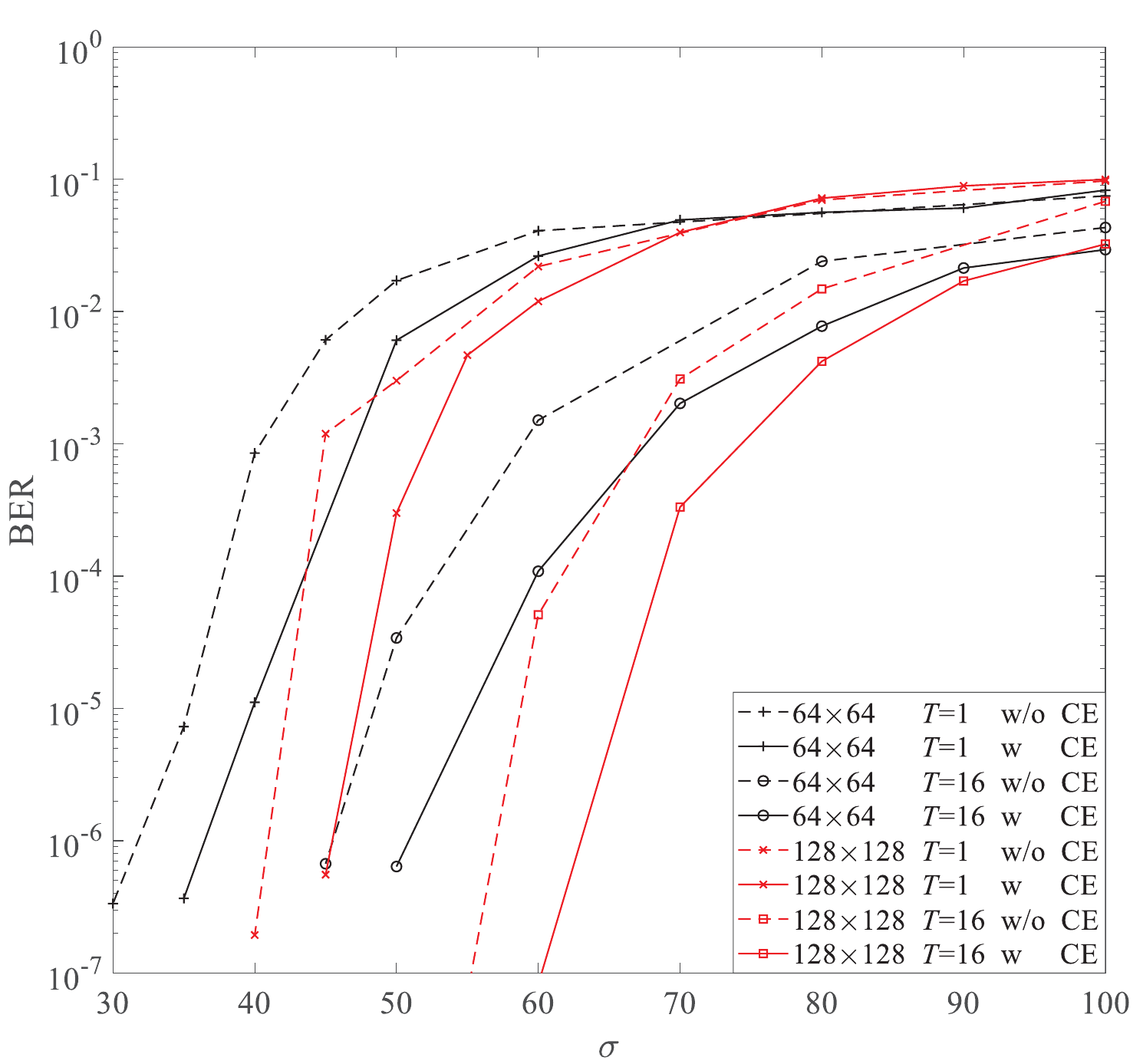}
\centering
\caption{BER comparison between the IRA-coded ReRAM channel with and without channel estimation (CE), where  memory array size is $m\times n=64\times64, 128\times128$ and joint-coding array number is $T=1, 16$.} \label{fig:BER_channel_est}
\end{figure}

\begin{figure}[t]
\includegraphics[width=
3.5 in]{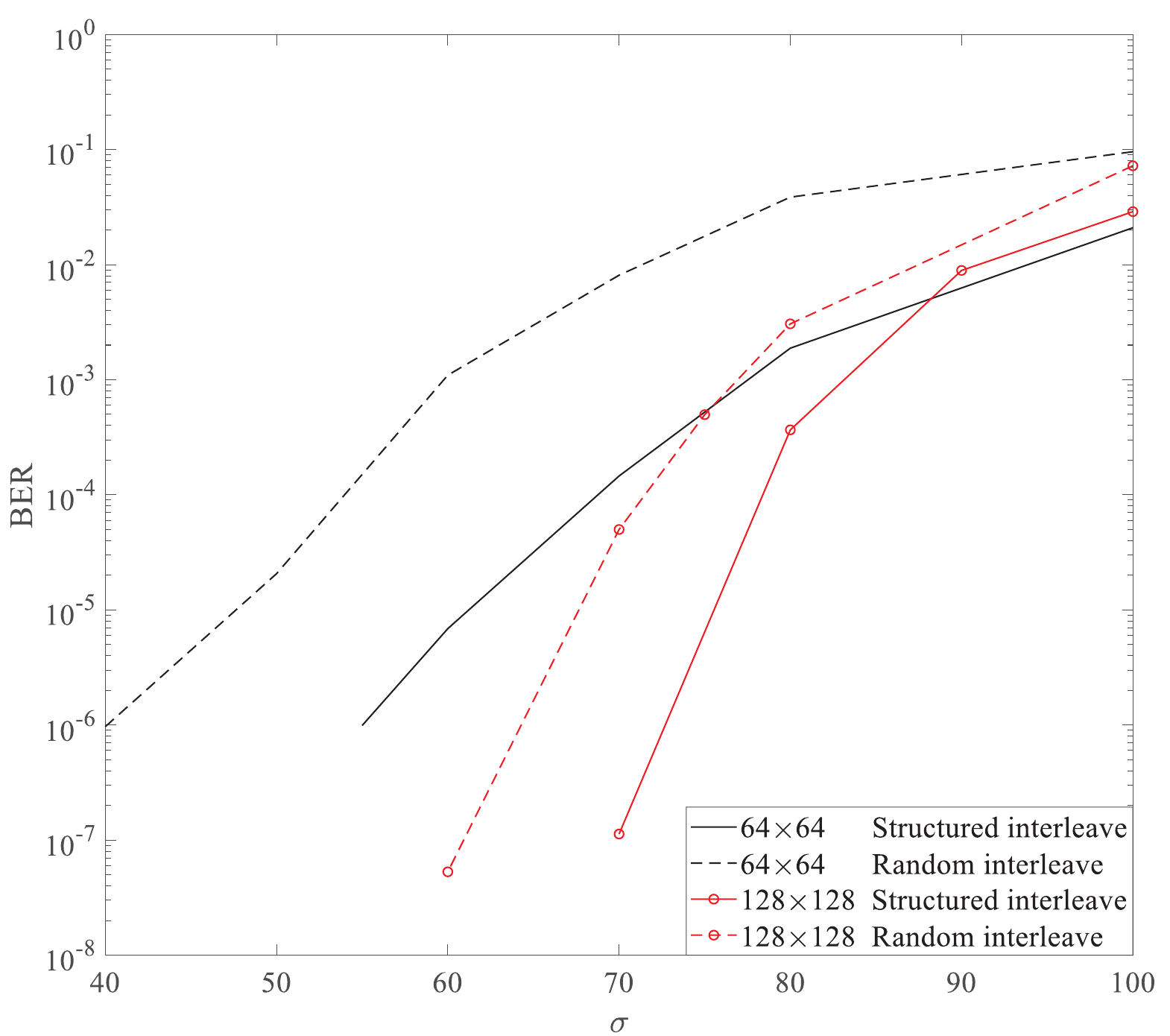}
\centering
\caption{BER comparison between our proposed structured interleaving and random interleaving over the i.i.d. $(\epsilon_q, \sigma)$-channel.} \label{fig:interleave}
\end{figure}

For the ECC, we adopt an IRA code which is a type of irregular low-density parity-check (LDPC) code that is able to approach the i.i.d. channel capacity with low encoding and decoding complexity \cite{RA,IRA}. We consider the code design for two ReRAM systems with a memory array sizes of $m\times n=64\times64$ and $128\times128$, and at a noise level of $\sigma=100$.  We adopt the data shapers with a $(L=4)$-repeater and 4-to-1 mappers with Mappings A and B (Fig.~\ref{fig:Mapping4}). Mappings A and B produce data distributions of $q=\frac{5}{16}$ and $q=\frac{3}{16}$, respectively, which approach the maximum storage efficiency for the considered ReRAM systems  (Fig.~\ref{fig:capacity}). Here the maximum storage efficiencies for $m\times n=64\times64$ and $128\times128$ ReRAM systems are $\max_qC_q(\epsilon_q, \sigma=100)=0.660$ and $0.494$ bit/cell, respectively.  We optimize the IRA code for these two cases over an i.i.d. $(\epsilon_q, \sigma)$-channel using a density evolution method \cite{LDPC_DE}. The code parameters for these two ReRAM systems are listed in TABLE~\ref{tab:code}. Our codes achieve $R=0.542824$ and $0.414735$ bit/cell, which are close to the capacity bound with gaps of about $0.12$  and $0.08$  bit/cell.

In Figs.~\ref{fig:BER64} and \ref{fig:BER128}, we simulate the BER of the two coded ReRAM systems in TABLE~\ref{tab:code} with the across-array coding strategy over both the ReRAM channels (solid lines) and the equivalent block-varying $(\fat{\epsilon}, \sigma)$-channels (labeled by $\circ$). Message-passing decoding were employed for both the de-shaper and the ECC decoder, where the decoding of IRA codes can be found \cite{IRA}.  For the $m\times n=64\times 64$ system, we employ a code length $N=64\times64=4096$ and for $128\times128$, we adopt $N=128\times128=16384$. In both figures, the BERs over the ReRAM channels are almost the same as those over the block-varying $(\fat{\epsilon}, \sigma)$-channels, thus demonstrating the proposed channel equivalence. The BERs improve as $T$ increases and approach the decoding performance over the i.i.d. $(\epsilon_q, \sigma)$-channel, which is the performance limit for ReRAM channels as $T\rightarrow\infty$. This performance improvement as $T$ increases can be regarded as a ``diversity" gain by assigning the codeword to multiple memory arrays.  For comparison, we also illustrate the BER performances of the same IRA codes without data shaping $(q=1/2)$ over i.i.d. $(\epsilon_q, \sigma)$-channel. The decoding performance deteriorates significantly due to the lack of data shaping.

To emphasize the importance of the proposed real-time channel estimation, in Fig.~\ref{fig:BER_channel_est}, we compared the BERs between the IRA-coded ReRAM channels with and without channel estimation for memory array sizes $m\times n=64\times64, 128\times128$ and  joint-coding array numbers $T=1, 16$. We employ the same code parameters listed in TABLE~\ref{tab:code} for both cases. For the IRA-coded ReRAM channel without channel estimation, we employ the average channel parameter $\epsilon_q$ for the decoder. We observe that by applying the channel estimation, the BER improvement is obvious for each comparison pair.

Note that the codes in TABLE~\ref{tab:code} are designed at $\sigma=100$ for the i.i.d. $(\epsilon_q, \sigma)$-channel, which means that theoretically the code  can be decoded without errors at $\sigma=100$. However, the actual decoding performances shown in Figs.~\ref{fig:BER64} and  \ref{fig:BER128} are much worse. This is because with density evolution, the code is designed under an infinite code length assumption, while in our simulations finite-length codes were employed. In other words, the codes are asymptotically decodable at $\sigma=100$ as the code length approaches infinity. For the same reason, the $128\times128$ ReRAM system achieves a lower BER performance than the $64\times64$ system since a much longer code is employed in the $128\times128$ system although the codes in both systems are designed at the same noise level of $\sigma=100$.

In Fig.~\ref{fig:interleave}, we provide BER comparisons between our proposed structured interleaving scheme and the random interleaving for the codes in TABLE~\ref{tab:code} over the i.i.d. $(\epsilon_q, \sigma)$-channel.  For both codes, our proposed structured interleaving scheme outperforms the random interleaving. The BER curves are steeper with structured interleaving than that with random interleaving. This verifies our analysis in Section~\ref{sec:shaper}. Note that similar performance gain can also be observed from the ReRAM channels, which are the block-varying forms of the $(\epsilon, \sigma)$-channel.

\section{Conclusion} \label{sec:conclusion}
 We have considered the design of effective channel coding
schemes to tackle both the sneak-path interference and the
additive noise for the ReRAM channels. We have proposed
an across-array coding strategy to mitigate the channel instability. It also enables
a ``diversity" gain during decoding. By employing TIN decoding, the ReRAM channel is equivalent to a block-varying channel whose status is not data-dependent, based on which, we proposed the capacity limit as well as a coding scheme. We have also proposed a
real-time channel estimation scheme to obtain the sneak-path
rates of the $T$ arrays, based on which an ESE calculates the
LLR for each coded bit for decoding.
To deal with the channel asymmetry, we proposed an ECC concatenated with a data shaper, where the later forms the desired input data distribution to achieve the maximum information rate.  With an optimal ECC design, the ReRAM system achieved a high storage efficiency with a gap of less than 0.1 bit/cell from the ReRAM channel capacity limit.

We would also like to point out some possible extensions that lead to our future work. Although we only considered the AWGN noise in this paper, our work can be easily extended to other types of noises, such as the lognormal noise, through reformulating the channel capacity and the LLR formula in the ESE. Moreover, to consider more general sneak-path models that involve  multiple sneak paths affecting a read cell,  the channel model in Fig.~\ref{fig:ESchannel} should be modified as a binary input multi-level output channel, where the types of the output signal levels depend on the corresponding sneak-path combinations \cite{Ben}.



\begin{thebibliography}{1}

\bibitem{Strukov} D. B. Strukov, G. S. Snider, D. R. Stewart, and R. S. Williams, ``The
missing memristor found," Nature, vol. 453, no. 7191, p. 80, 2008.

\bibitem{Zidan} M. A. Zidan, H. A. H. Fahmy, M. M. Hussain, and K. N. Salama,
``Memristor-based memory: The sneak paths problem and solutions,"
Microelectron. J., vol. 44, no. 2, pp. 176--183, 2013.

\bibitem{Yuval}
Y. Cassuto, S. Kvatinsky, and E. Yaakobi, ``Information-theoretic sneakpath mitigation in memristor crossbar arrays," \emph{IEEE Trans. Inf. Theory.}, vol. 62, no. 9, pp. 4801--4813, Sep. 2016.


\bibitem{Ben} Y. Ben-Hur and Y. Cassuto, ``Detection and coding schemes for sneakpath interference in resistive memory arrays," \emph{IEEE Trans.
Commun.}, vol. 67, no. 6, pp. 3821--3833, Feb. 2019.

\bibitem{CZH} Z. Chen, C. Schoeny, and L. Dolecek, ``Pilot assisted adaptive thresholding for
sneak-path mitigation in resistive memories with
failed selection devices," \emph{IEEE Trans.
Commun.}, vol. 68, no. 1, pp. 66--81, Jan. 2020.

%
%

\bibitem{Huang} P.~Huang, P.~H.~Siegel, and E. Yaakobi, ``Performance of multilevel flash memories with different binary labelings: a multi-user perspective," \emph{IEEE JSAC}, vol. 34, no. 9, pp.2336--2353, Sep. 2016.

\bibitem{Raymond} R. Knopp and P. A. Humblet, ``On coding for block fading channels," \emph{IEEE Trans. Inf. Theory}, vol. 46, no. 1, pp.189--205, Jan. 2000.

\bibitem{Vontobel}P. O. Vontobel,W. Robinett, P. J. Kuekes, D. R. Stewart, J. Straznicky, and
R. S.Williams, ``Writing to and reading from a nano-scale crossbar memory
based on memristors," \emph{Nanotechnology}, vol.~20, no.~42, Oct. 2009.

\bibitem{Zhou}J. Zhou, K. Kim, and W. Lu, ``Crossbar RRAM arrays: selector device requirements during read operation," \emph{IEEE Trans. Electron Devices}, vol. 61, no. 5, pp. 1369--1376, May 2014.

\bibitem{Luo} Q. Luo, X. Xu, T. Gong, and H. Lv, ``8-layers 3D vertical RRAM with excellent scalability towards storage class memory applications," \emph{in Proc. IEEE IEDM 2017}, pp. 2.7.1--2.7.4.
    
    \bibitem{Mceliece} R. J. Mceliece and W. E. Stark, ``Channels with block interference," \emph{IEEE Trans. Inf. Theory}, vol. IT-30, no. 1, pp.44--53, Jan. 1984.

\bibitem{RA}D. Divsalar, H. Jin, And R. J. Mceliece. ``Coding theorems for `turbo-like' codes." in \emph{Proc. 36Th Allerton Conf. On Communication, Control And Computing}, Allerton, Illinois, Sept. 1998, pp. 201--210.

\bibitem{IRA} S.~ten Brink and G.~Kramer, ``Design of repeat-accumulate codes
for iterative detection and decoding," \emph{IEEE Trans. Signal
Processing}, vol.~51, no.~11, pp.~2764--2772, Nov.~2003.


\bibitem{LDPC_DE} T.J. Richardson and R.L. Urbanke, ``The capacity of low-density parity-check codes under message-passing decoding," \emph{IEEE Trans. Inf. Theory}, vol.~47, no.~2, pp.~599 -- 618, Feb. 2001.


%
%

\end{thebibliography}
\end{document}